\documentclass[review,jog]{igs}

\usepackage[utf8]{inputenc}
 \usepackage{url}
 \usepackage{amsmath}
\usepackage{empheq}
\usepackage{mathabx}
\usepackage{graphicx}
\usepackage{siunitx}
\usepackage{algorithm}
\usepackage{algpseudocode}
\usepackage{amssymb}
\usepackage{enumitem}
\usepackage{amsmath}
\usepackage[nopatch]{microtype}
\usepackage{booktabs}
\usepackage{enumitem}
\usepackage{rotating}
\usepackage{booktabs}
\usepackage[utf8]{inputenc}
\usepackage{enumerate}
\usepackage{subcaption}
\usepackage{pythonhighlight}
\usepackage{epsfig}
\usepackage{amsmath}
\usepackage{mathrsfs}  
\usepackage{amsthm}

\usepackage{graphicx}
\usepackage{float}
\usepackage{bm}
\usepackage{graphicx}
\usepackage{multirow}
\usepackage{pdflscape}

\renewcommand{\d}{\mathrm{d}}

\def\tsc#1{\csdef{#1}{\textsc{\lowercase{#1}}\xspace}}
\tsc{WGM}
\tsc{QE}
\tsc{EP}
\tsc{PMS}
\tsc{BEC}
\tsc{DE}
\nolinenumbers
\jourvolume{XX}
\jourissue{X}
\jourpubyear{20XX}

\begin{document}


\title[A vertically integrated model with phase change for aquifers in cold firn]{A vertically integrated model with phase change for aquifers in cold firn}
\author[Shadab and others]{Mohammad Afzal Shadab$^{1,2,\dagger}$, Howard A. Stone$^{3}$, and Reed M. Maxwell$^{1,2,4}$}

\affiliation{
$^{1}$ Department of Civil and Environmental Engineering, Princeton University, Princeton NJ 08544 \\
$^{2}$ Integrated GroundWater Modeling Center, Princeton University, Princeton NJ 08544 \\
$^{3}$ Department of Mechanical and Aerospace Engineering, Princeton University, Princeton NJ  08544 \\
$^{4}$ High Meadows Environmental Institute, Princeton University, Princeton NJ 08544 \\
Correspondence: Mohammad Afzal Shadab
\email{mashadab@princeton.edu}}

\begin{frontmatter}

\maketitle

\begin{abstract}
Surface meltwater from glaciers and ice sheets contributes significantly to sea-level rise, yet the processes governing its transport and retention within cold firn remain poorly constrained, particularly in multiple dimensions. Here we present a multidimensional, vertically integrated modeling framework for aquifers in cold firn that incorporates phase change and residual trapping of liquid water. This mathematical framework, together with its numerical implementation, extends terrestrial groundwater models to describe aquifers expanding within otherwise cold firn, highlighting the analogous physics governing both systems. We derive semi-analytical solutions for finite-volume aquifers and validate them against numerical simulations and higher-fidelity model results. These solutions elucidate key features of meltwater dynamics and provide benchmarks for firn hydrologic models. We further demonstrate the three-dimensional expansion of an aquifer in cold, heterogeneous firn. Both the semi-analytical and numerical results show that lateral aquifer propagation slows at lower initial firn temperatures due to porosity reduction and associated loss of liquid water from freezing. Overall, this framework provides new insights into the formation and expansion of firn aquifers in percolation zones and helps clarify how subsurface meltwater storage modulates meltwater fluxes, surface mass loss, and contributes to global sea-level rise.


\end{abstract}

\end{frontmatter}

\section{  Introduction}\label{sec1}
Understanding the subsurface migration and retention of meltwater in sintered and compacted snow, called firn, on ice sheets is a central challenge in cryospheric hydrology \citep{firn2024firn}. Surface meltwater generated on glaciers and ice sheets may percolate into the porous firn layer, where it can be stored as a liquid, refreeze, or propagate laterally as subsurface flow, thereby modulating the amount of water that eventually contributes to runoff and sea level rise \citep{harper2012greenland,Meyer2017,firn2024firn}. In cold conditions, the interplay between percolation and freezing alters pore-space structure, controlling meltwater pathways and storage \citep{Colbeck1972, marsh1984wetting, Pfeffer1996}. Classical firn hydrology experiments describe both the advancement of a uniform wetting front and preferential flow through discrete channels or pipes, highlighting the sensitivity of infiltration to meltwater flux, initial temperature, density, and grain size \citep{williams2010visualizing,humphrey2012thermal, Avanzi2016}. Dye-tracing experiments have confirmed the occurrence of lateral and preferential flow within cold firn and snow, indicating complex three-dimensional meltwater redistribution \citep{marsh1984wetting,pfeffer1998formation,harper2012greenland,humphrey2012thermal,wever2016simulating,clerx2022situ,Moure2023,Jones2024}. Thermal processes further modulate these flows as freezing releases latent heat, warms the firn, and reduces porosity, often creating ice layers that impede subsequent infiltration \citep{Humphrey2021,shadab2024mechanism}. Modern multilayer firn hydrologic models simulate vertical, one-dimensional meltwater percolation coupled with thermodynamics and freezing \citep{Steger2017,Meyer2017,Vandecrux2020,Ashmore2020,Moure2023,gardner2023glacier,shadab2024mechanism,shadab2025unified,jones2026influence}, but they typically neglect lateral transport. Thus, a theoretical framework to describe lateral flow embedded within a thermally evolving, partially freezing porous medium remains underdeveloped, particularly for expanding firn aquifers where heat transport, phase change, and pore closure modulate flow dynamics. In this work, we develop mathematical and numerical frameworks that contribute to an improved understanding of the dynamics of firn aquifers.

In cold firn, a portion of infiltrating liquid water may freeze within the pore space, reducing permeability and inhibiting further flow \citep{colbeck1976analysis,Clark2017,Meyer2017,Jones2024,
shadab2024mechanism,shadab2025unified}. These studies highlight the strong coupling between heat transfer, phase change, and fluid flow that governs the evolution of permeability and saturation in cold porous media. When lateral hydraulic gradients are present, water may spread laterally within a saturated region, behaving as a firn aquifer within the porous medium. 
In the firn hydrology research community, firn aquifers have drawn increasing attention as they act to buffer meltwater input and modulate runoff timing and magnitude. Some observations of firn aquifers in Antarctica, Greenland, and Svalbard, along with their typical geometries, are summarized in Table \ref{tab:firn_aquifer_obs}. Typically, firn aquifers are located tens of meters below the surface and are tens of meters thick. Their thickness changes over the seasons, and a higher meltwater supply increases their thickness. Figure~\ref{fig1:sample-aquifer}\textit{a} shows a firn aquifer in Antarctica, whereas Figure~\ref{fig1:sample-aquifer}\textit{b} illustrates the large extent of the perennial firn aquifer in Greenland. Warming will enhance melt production and accelerate the transition from accumulation to ablation, likely increasing their spatial extent over time \citep{miller2023hydrologic}. 

Many models exist for firn aquifer dynamics in the vertical direction that consider more detailed physics. For example, \cite{Meyer2017} include variably saturated flow hydrology, compaction, thermodynamics, and phase change \citep[also, see numerical models in][]{Steger2017,Vandecrux2020}. However, there are few studies on modeling firn aquifers in two and three dimensions, as summarized in Table \ref{tab:firn_modeling_summary}. These studies include vertically-integrated firn aquifer or high-fidelity firn hydrologic (SUTRA-2D) models that resolve the vertical flow. High fidelity models are computationally expensive and, in some cases, intractable; therefore, vertically integrated models are utilized, for example, in the groundwater science community \citep{Bear_1972, colbeck1978physical, pinder2006subsurface, shadab2024hyperbolic}. An extensive literature on aquifers in porous media offers useful foundations \citep{benjamin1968gravity,kochina1983groundwater,huppert1995gravity,simpson1999gravity,ungarish2009introduction,zheng2022influence}. Most existing firn aquifer models implement the terrestrial groundwater model for the aquifer, which is based on the vertical integration of liquid water conservation. However, to the best of our knowledge, existing firn hydrology models have not addressed lateral groundwater-style flow in cold firn with freezing, which reduces liquid water and produces changes in porosity and permeability. Although thermodynamics may have been coupled in vertically integrated models via 1D (vertical) firn hydrology, it has never been incorporated explicitly into the vertically integrated flow equations themselves. This represents a conceptual gap between terrestrial groundwater hydrology and the hydrology of firn aquifers, where heat transport and phase change, causing water loss and pore closure, could strongly influence aquifer dynamics. In this paper, we aim to address this conceptual gap by extending the concepts from terrestrial groundwater hydrology to model and study firn aquifers.

Here, we present a vertically integrated modeling framework for aquifers in cold firn that accounts for phase change due to freezing. 
The framework applies to regimes where gravitational and hydraulic forces dominate the capillary forces in driving the meltwater, and heat advection dominates heat conduction. 
The paper is organized as follows. In Section \ref{sec2:VIM_formulation}, we present the governing equations for aquifers in cold firn while incorporating the residual trapping of liquid water and porosity reduction due to freezing. In Section \ref{sec3:const-vol}, we present semi-analytical (self-similar) solutions for a finite volume firn aquifer in an otherwise uniform and cold firn, which are used for analysis and numerical validation against the firn aquifer model and a higher-fidelity model. In Section \ref{sec4:corr-rnd-field}, we simulate three-dimensional aquifer propagation in a cold firn with heterogeneous porosity and temperature fields, while in Section \ref{sec5:discussion-limitations} we discuss the model features and limitations. 

Our framework contributes to firn hydrology by providing a low dimensional but physics-based model for predicting the lateral redistribution of meltwater in cold regions. Because lateral propagation can significantly alter the spatial extent and residence time of subsurface meltwater, our results have implications for meltwater retention, freezing, and runoff partitioning. The proposed mathematical framework enables the development of fast and accurate numerical models to understand and investigate firn aquifer dynamics and provides benchmarks to verify and validate multidimensional firn hydrologic models. It encapsulates previous firn aquifer problems and may be particularly useful in regions experiencing the expansion of firn aquifers into cold regions.

\section{ Firn Aquifer Model formulation}\label{sec2:VIM_formulation}
\begin{figure}
    \centering
        \includegraphics[width=\linewidth]{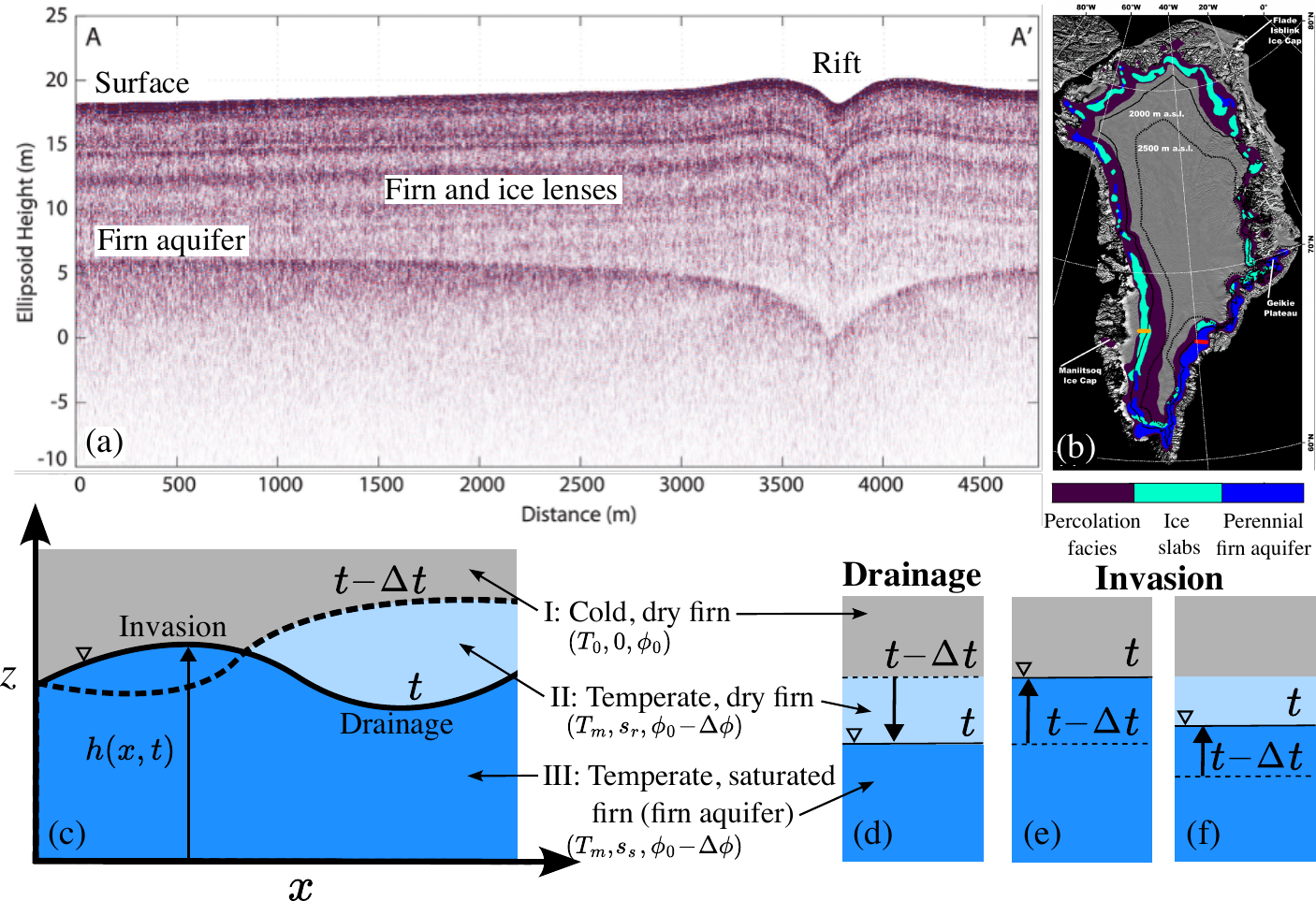}
    \caption{Aquifers in firn: (a) Ground penetrating radar (GPR) observations showing firn aquifers in Antarctica (modified from \cite{montgomery2020hydrologic}). (b) Map showing extent of perennial firn aquifers in Greenland (modified from \cite{miller2022empirical}). (c) Conceptual diagram showing aquifer in cold firn in two-dimensional cartesian ($x,z$) coordinates. The lateral coordinates could be $\tilde{\textbf{x}}\equiv(x,y)$ in cartesian coordinates and $\tilde{\textbf{x}}\equiv(r,\theta)$ in cylindrical coordinates with $z$ being the vertical coordinate. Representations of dynamics of the aquifer in firn during (d) drainage, (e) invasion in cold regions, and (f) invasion in warmed regions. Here the variables ($T,s_w,\phi$) denote the local temperature, saturation and porosity, respectively. The water table is located at $h(x,t)$ from the impermeable base at time $t$ and the dashed line shows the location of the water table at the previous time instant $t-\Delta t$. The domain is divided into three regions- I: cold and dry ($T_0,0,\phi_0$), II: temperate and dry ($T_m,s_r,\phi_0-\Delta \phi$), and III: temperate and saturated ($T_m,s_s,\phi_0-\Delta \phi $). }
    \label{fig1:sample-aquifer}
\end{figure}

We consider a cold firn with an initial vertically averaged temperature field $T_0(\tilde{\textbf{x}})$ and porosity field $\phi_0(\tilde{\textbf{x}})$ at lateral locations $\tilde{\textbf{x}}=(x,y)$ or $(r,\theta)$ (Figure~\ref{fig1:sample-aquifer}\textit{c}). We use a tilde to denote vectors in lateral dimensions while excluding the vertical coordinate $z$. As such, the initial temperature and porosity of the cold firn are only functions of horizontal spatial locations $\tilde{\textbf{x}}$ and not of the vertical location $z$.

\begin{landscape}
\begin{table}
\caption{A brief summary of observational studies of firn aquifers across Greenland, Antarctica, and Svalbard. Depth refers to the depth measured from the surface to the top of the water table (WTD) and thickness refers to the vertical extent of liquid water detected in the firn column.}
\begin{tabular}{@{} p{4cm} p{1.5cm} p{5.2cm} p{1cm} p{1.5cm} p{8cm} @{}}
\hline
\textbf{Reference} & \textbf{Location} & \textbf{Observation methods} & \textbf{Depth (m)} & \textbf{Thickness (m)} & \textbf{Remarks} \\
\hline
\cite{koenig2014initial} & Greenland & Borehole drilling, temperature and density logs, direct water recovery & 12--37 & 24.7 & In-situ confirmation of a perennial firn aquifer with year-round liquid water at 0\,°C. \\

\cite{christianson2015dynamic} & Svalbard & Ice-penetrating radar and GPS & 0-25 & 10--20 & WTD changes between -3.5 and 3.5 m in 2005-07. \\

\cite{forster2014extensive} & Greenland & Ground and airborne radar & 5--50 & -- & Radar-based mapping of extensive firn aquifer band. \\

\cite{miege2016spatial} & Greenland & Airborne radar, ground based radar & 4--40 & -- & Aquifer region coincides with high accumulation and melt rates and WTD changes by 1.1 m in 2013. \\

\cite{montgomery2017investigation} & Greenland & Ground penetrating radar (GPR), seismic refraction experiments & 27.7 & 11.5 & Aquifer structure shows that water storage is controlled by accumulation and permeability contrasts. \\

\cite{miller2018direct} & Greenland & Borehole hydraulic tests, tracers, temperature profiles & 25--30 & 10--20 & Direct evidence of lateral saturated flow with estimated hydraulic conductivity $K \approx 10^{-6}$--$10^{-5}$ m\,s$^{-1}$. \\

\cite{chu2018retrieval} & Greenland & Airborne radar, signal attenuation inversion & 15--27 & 6--17 & Firn aquifers are dynamically sensitive to annual changes in surface forcing and WTD changed about 2-7.5 m/yr in 2012-14. \\

\cite{legchenko2018estimating} & Greenland & Magnetic Resonance Sounding (MRS) and GPR & $\sim$18--21 & 11--16 & 
The volume of water stored in their study area increases by about 36\% in about a year (i.e., 2015-16). \\

\cite{miller2018direct} & Greenland & Borehole observations & 10--20 & 8--37.8 & Aquifer behaves as a dynamic groundwater system with observed seasonal fluctuations in head. \\

\cite{killingbeck2020integrated} & Greenland & Borehole logs, radar, seismic velocities & 10--20 & 8--18 & Refrozen ice layers within the aquifer do not appear to impede lateral flow but could reduce vertical flow.  \\

\cite{cicero2023firn} & Greenland & Field observations, crevasse discharge mapping & 5--40 & -- & Firn aquifer discharges into crevasses; links surface melt to englacial and subglacial drainage pathways. \\

\cite{montgomery2020hydrologic} & Antarctica & Boreholes, density, hydraulic tests & 13.39--13.46 & 16.2 & Aquifer has high permeability, sustained liquid storage, and is draining laterally into a large nearby rift. \\

\cite{van2025long} & Svalbard & Boreholes, GPR & 10-40 & -- & Persistent aquifer observed in 2017--19 and model shows evolution under changing climate. \\
\hline
\end{tabular}
\label{tab:firn_aquifer_obs}
\end{table}
\end{landscape}

\begin{landscape}
\begin{table}
\caption{A summary of multidimensional firn aquifer models.}
\renewcommand{\arraystretch}{1.25}
\begin{tabular}{p{2cm} p{2.5cm} p{2cm} p{2cm} p{3cm} p{2.5cm} p{6cm}}
\hline
\textbf{Reference} & \textbf{Framework} & \textbf{Dimensions} & \textbf{Time-dependent} & \textbf{Representation of firn aquifer} & \textbf{Heat or Phase change} & \textbf{Key advances or limitations} \\
\hline
\cite{Miege2016} & Seep2D & 2D & No & Vertically averaged & No & Empirical 2D reconstruction of aquifer geometry from GPR data with no explicit thermodynamics modeled. \\
\hline
\cite{miller2023hydrologic} & SUTRA-ICE (variably saturated, energy-coupled) & 2D & Yes & Fully resolved & Yes (high fidelity) & Captures freezing fronts, simulates simple geometry, and is computationally expensive. \\
\hline
\cite{clerx2024modelling} & Physics-based quasi-2D model + superimposed ice formation model & 2D & Yes & Darcy's law used pointwise between 1D columns  & Yes (heat conduction loss forms superimposed ice) & Simulates lateral runoff over ice slabs and shows aquifer thins due to ice formation. Model does not solve vertically integrated model for lateral flow, thus, total liquid water mass (liquid + solid ice) may not be conserved.  \\
\hline
\cite{van2025long} & MODFLOW-6 + firn model & 3D & Yes & Vertically integrated & Partial (via firn coupling) & The aquifer model does not explicitly include meltwater freezing due to heat loss. Heat transport is restricted to the 1D bucket-hydrology model above the firn aquifer. \\
\hline
{Present work} & Firn aquifer model with phase change & {2D/3D} & Yes & {Vertically integrated} & {Yes (heat advection included in governing equation)} & {Includes heat advection caused freezing and can be coupled with 1D firn models. Verified and validated against analytical and high-fidelity numerical solutions.} \\
\hline
\end{tabular}
\label{tab:firn_modeling_summary}
\end{table}
\end{landscape}
We now assume an unconfined aquifer over a horizontal base (ice slab or bedrock) within the firn with a high aspect ratio (length/height $\gg 1$), such that the velocities in the vertical direction ($z$) are negligible compared to lateral velocities for most of the current, and the pressure in the aquifer is hydrostatic. Air is assumed not to participate in hydrology or thermodynamics due to its low density, viscosity, and thermal conductivity. We also assume that there is no loss of heat from the base of the aquifer.

Assuming local thermodynamic equilibrium and neglecting the effect of heat conduction compared to heat advection, a simplified model for a representative elemental volume (REV) involves the freezing of liquid water while accounting for density differences when it invades a new pore space to warm the solid ice grains to the melting temperature. This idea of freezing has been implemented previously for 1D (vertical) infiltration in firn \citep{colbeck1976analysis,Clark2017,shadab2025unified} but not for lateral propagation via firn aquifers. During invasion, the freezing leads to the partial closure of the pore space, which reduces the porosity by $\Delta \phi(\tilde{\textbf{x}})$ and causes the loss of liquid water that affects the dynamics of the firn aquifer. As such, the aquifer always experiences reduced porosity $\phi'(\tilde{\textbf{x}})=\phi_0(\tilde{\textbf{x}})-\Delta \phi(\tilde{\textbf{x}})$ and is at the melting temperature. However, during drainage, a volume fraction of liquid water $\phi'(\tilde{\textbf{x}}) s_r$ is retained due to the trapping of residual water in the pore space, with $s_r$ being the residual saturation.

The resulting porous medium can be divided into three regions (see Figures~\ref{fig1:sample-aquifer}\textit{c}-\textit{f}). These regions arise from the thermal and hydrological history of the porous medium. Region I represents undisturbed cold, dry firn. When liquid water invades Region I, freezing occurs to warm the surrounding firn to the melting temperature, creating Region III with reduced porosity $\phi_0 - \Delta \phi$ and full saturation $s_s$. When the aquifer drains from a region, it creates Region II with reduced porosity and residual trapped liquid water. More specifically, Region I is cold and dry, with temperature $T=T_0$, water saturation $s_w=0$, and porosity $\phi = \phi_0$, which we write together as a state variable ($T,s_w,\phi)=(T_0,0,\phi_0$). Region II is a temperate and (nearly) dry region that has been invaded by liquid water in its history, but the aquifer has drained out of it. The resulting porosity has been reduced from $\phi_0$ to $\phi'=\phi_0 - \Delta \phi(\tilde{\textbf{x}})$ due to meltwater freezing, and is at residual water saturation $s_r$ and melting temperature $T_m$. Thus, Region II is defined by the state variables ($T_m,s_r,\phi_0-\Delta \phi$). Region III forms the unconfined aquifer that is temperate ($T=T_m$) where $T_m$ being the melting temperature, saturated ($s_w=s_s$), and has experienced meltwater freezing leading to a porosity of $\phi_0-\Delta \phi$. Thus, the state variables in region III are ($T_m,s_s,\phi_0-\Delta \phi $). The height of the unconfined aquifer $h(\tilde{\textbf{x}},t)$ in Region III is governed by a vertically integrated model for the evolution of liquid water, which is derived from the enthalpy conservation (see Appendix \ref{appB_enthalpy}) or total water conservation (see Appendix \ref{appsec:composition}) as

\begin{align}
\phi'  s_s \, \frac{\partial h}{\partial t} + &R \, \frac{\partial h}{\partial t} -  \nabla \cdot \left( K h \nabla h \right)  = 0,\label{eq:governing2}
\end{align}
where
\begin{align}
 R = \begin{cases}
  \Delta \phi(T_0,\phi_0)\rho_i/\rho_w, \quad &\text{for } \frac{\partial h}{\partial t} > 0 \quad \text{and} \quad  h(\tilde{\textbf{x}},t)=h_{max}[\tilde{\textbf{x}},t]\\
  -\phi' s_r, \quad &\textrm{otherwise}.
 \end{cases}\label{eq:governing2extra}
\end{align}
Here the gradient and divergence operators exist only in lateral directions, such as cartesian coordinates where $\tilde{\textbf{x}}\equiv(x,y)$ or cylindrical coordinates where $\tilde{\textbf{x}}\equiv(r, \theta)$. Furthermore, $\rho_i$ and $\rho_w$ are the densities of the ice and liquid water phases (kg/m$^3$) respectively, $K$ is the saturated hydraulic conductivity defined as $K=\frac{k(\phi')\Delta  \rho g}{\mu}$, $\phi'=\phi_0 - \Delta \phi(\tilde{\textbf{x}})$ is the reduced porosity that the aquifer experiences following the instantaneous phase change during initial invasion, $\Delta \rho$ is the density difference between water and air (kg/m$^3$), $k$ is the permeability of the porous media (m$^2$), $\mu$ is the dynamic viscosity of liquid water at the melting temperature (Pa$\cdot$s), and $g$ is the acceleration due to gravity (m/s$^2$). 

In the governing equation \eqref{eq:governing2}, the first term ($\phi'  s_s \, {\partial h}/{\partial t}$) represents water accumulation in the saturated aquifer and the third term ($-  \nabla \cdot \left( K h \nabla h \right)$) describes lateral transport due to hydraulic gradients. Both of these terms exist in prior firn aquifer models that are based on terrestrial groundwater models. However, the second term ($R \, {\partial h}/{\partial t}$) is new and accounts for liquid water loss due to freezing $R=\Delta \phi(T_0,\phi_0)\rho_i/\rho_w$ (Equation~\ref{eq:governing2extra}), causing instantaneous warming of the surrounding cold firn to melting temperature when the aquifer invades the cold region, given by the condition $\frac{\partial h}{\partial t} > 0 \quad \text{and} \quad  h(\tilde{\textbf{x}},t)=h_{max}[\tilde{\textbf{x}},t]$ (see Figure~\ref{fig1:sample-aquifer}\textit{e}); otherwise, this new term corresponds to the warmed firn with residual liquid water trapped in the pore space, given by $R = -\phi' s_r$ (see Figures~\ref{fig1:sample-aquifer}\textit{d, f}). Here the vertical height $h_{max}$ is the largest thickness that the aquifer has achieved at a particular location $\tilde{\textbf{x}}$ up to the time of consideration $t_1$ described mathematically as

\begin{align}\label{eq:hmax}
    h_{max}[\tilde{\textbf{x}},t_1] = \max_{t\leq t_1}(h[\tilde{\textbf{x}},t]) \quad \forall \tilde{\textbf{x}}.
\end{align}

Furthermore, the reduction in porosity $\Delta \phi$, due to the freezing of liquid water ($\Delta \phi \rho_i/\rho_w$) that warms the surrounding firn, follows from the energy balance as

\begin{align}
    \Delta \phi(T_0(\tilde{\textbf{x}}),\phi_0(\tilde{\textbf{x}})):=  \frac{ c_{p,i} (T_m-T_0(\tilde{\textbf{x}})) (1-\phi_0(\tilde{\textbf{x}}))}{L} \label{eq:energy-balance}
\end{align}
while accounting for the density change during freezing. Here $L$ is the latent heat of fusion of water (J/kg), $c_{p,i}$ is the specific heat of ice at constant temperature (J/kg$\,^\circ$C), and $1-\phi_0(\tilde{\textbf{x}})=\phi_i(\tilde{\textbf{x}})$ is the volume fraction of ice. There could be a case where the freezing caused by heat advection can lead to the formation of an impermeable ice layer when $\Delta \phi = \phi_c-\phi_0$ where $\phi_c$ is the cut-off porosity with zero hydraulic conductivity. The cut-off porosity is typically $\phi_c\sim 0.094$ corresponding to a density of 830 kg/m$^3$ \citep{cuffey2010physics}. This condition of freezing causing complete closure of pore space has been previously considered during infiltration process in \cite{colbeck1976analysis,Humphrey2021,shadab2025unified}. Equations (\ref{eq:governing2}-\ref{eq:governing2extra}) summarize the general dynamics of the firn aquifer on a horizontal base, including freezing and residual liquid water trapping. Furthermore, terms such as recharge, which causes an increase in aquifer height, and conduction, which causes a loss of liquid water, could be accounted for as source terms. 

We numerically solve the governing equation (\ref{eq:governing2}) along with the conditional statement (\ref{eq:governing2extra}) in one and two dimensions and refer to these solutions as ``numerical results'' throughout the manuscript unless stated otherwise. The partial differential equation (\ref{eq:governing2}) is solved using a conservative finite-difference framework called the discrete operator toolbox. For a pedagogical introduction to the discrete-operator toolbox, including the numerical implementation of gradient and divergence operators as well as boundary conditions, please see \cite{shadab2024hyperbolic}. The numerical model is second-order accurate in space and first-order accurate in time, and is based on explicit Euler time marching. In the next section, we demonstrate the utility of this mathematical framework and numerical model for two- and three-dimensional problems involving the expansion of cold regions. Using Equations (\ref{eq:governing2}-\ref{eq:governing2extra}), we will first develop theoretical solutions for aquifers expanding in cold firn that can help verify and validate vertically-integrated firn aquifer models or higher-fidelity, multidimensional firn hydrologic models. To clearly show the effect of freezing, we set the saturation at the saturated state $s_s=1$ and residual saturation $s_r=0$ in numerical solutions, but the residual saturation observed in drainage experiments is $s_r\approx 0.07$ \citep{coleou1998irreducible}. The hydraulic conductivity of the refrozen firn, $K(\phi')$, given in Equation~\eqref{eq:governing2}, is assumed to follow a power-law relationship between permeability and porosity, $k(\phi') = k_0 \phi'^3$, with $k_0 = 5.6 \times 10^{-11}$~m$^2$ \citep{Meyer2017,shadab2024mechanism,shadab2025unified}.

\section{ Solutions for expansion of a finite volume aquifer in cold firn}\label{sec3:const-vol}
\subsection{ Analytical solutions in cartesian and cylindrical coordinates}\label{sec:solutions}
\begin{figure}
    \centering
    \includegraphics[width=0.8\linewidth]{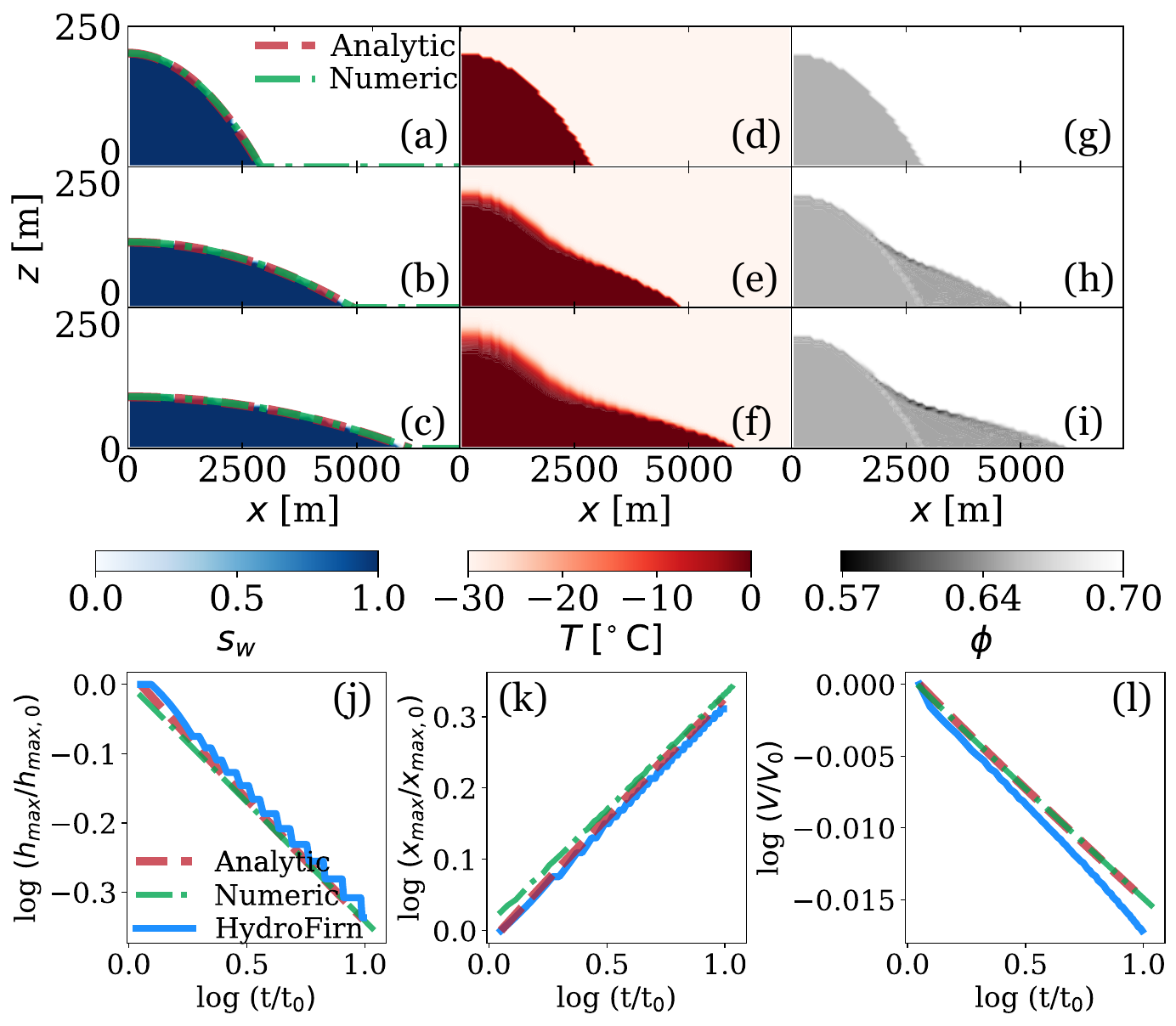}
    \caption{Expansion of an aquifer in an otherwise uniform, cold firn outside the aquifer with initial temperature $T_0 =-30\,^\circ$C and porosity $\phi_0=0.7$. Solutions of the aquifer height or hydraulic head at (a) $t=1$ year (initial condition), (b) $t=5$ years, and (c) $t=10$ years from theoretical solutions in cartesian coordinates (Analytic) and the numerical solutions (Numeric) of the vertically integrated model. The contour plots are the solutions from the higher-fidelity HydroFirn model for (a-c) saturation $s_w$, (d-f) temperature $T$, and (g-i) porosity $\phi$ at $t=$ 1, 5 and 10 years. Evolution of dimensionless (j) maximum height, (k) maximum length, and (l) volume of liquid water in the aquifer, scaled with respect their initial values at time $t_0=1$ year (subscript $0$ refers to the initial values). Semi-analytical and numerical solutions results agree with each other as well as HydroFirn model. Logarithms with base 10 ($\log_{10}$) are used in panels \textit{j} through \textit{l}. Supplementary video S1 shows the expansion of firn aquifer from the HydroFirn model (contour plots or solid blue lines) and semi-analytic solutions (red dashed lines).}
    \label{fig2:analyvsnum}
\end{figure}

    We consider a finite volume aquifer at $T_m=0\,^\circ$C in an otherwise cold porous medium at a uniform initial temperature $T_0 \leq T_m$ and porosity $\phi_0$ outside the firn aquifer (Figures~\ref{fig2:analyvsnum}\textit{a}, \textit{d}, \textit{e}). Initially, the medium is uniform with two discontinuous states: within the aquifer (Region III: $T=0\,^\circ$C, $s_w=s_s$, $\phi=\phi'=\phi_0-\Delta \phi$) and outside of the aquifer (Region I: $T=T_0$, $s_w=0$, $\phi=\phi_0$). The initial $\Delta \phi$ is chosen to be the pre-calculated value from energy balance \eqref{eq:energy-balance} and is identical to the freezing that will occur once the aquifer expands. Alternatively, the domain could be considered homogeneous ($T=T_0,s_w=0,\phi=\phi_0$ everywhere) if there had been no liquid water in the domain. As a result, in this case, the reduction in porosity due to the freezing of liquid water $\Delta \phi(\phi_0,T_0)$ is constant rather than a field, and the temperature rises to 0$\,^\circ$C in the regions where the firn aquifer is/was present. For this case, the height either monotonically decreases or increases with time. This further simplifies the condition for $R$ in Equation \eqref{eq:governing2extra} to $\partial h/\partial t>0$ for the invasion of liquid water into cold regions causing its partial freezing and $\partial h/\partial t \leq 0$ for aquifer drainage with residual liquid water trapping. Thus, the governing equations (\ref{eq:governing2}-\ref{eq:governing2extra}) can be rewritten in the following form in 1D cartesian ($x$) and cylindrical ($r$) coordinates, respectively, as

\begin{align} \label{eq:governing}
    \frac{\partial h}{ \partial t}=
    {\kappa^*} \frac{\partial}{\partial x}\left(\frac{\partial h^2}{\partial x} \right)
     \quad \text{and} \quad   \frac{\partial h}{ \partial t}=
    \frac{\kappa^*}{r} \frac{\partial}{\partial r}\left(r\frac{\partial h^2}{\partial r} \right),
\end{align}

where the coefficient $\kappa^*$ is

\begin{align}\label{eq:condition-Kochina}
    \kappa^*=
\begin{cases}
\kappa = \frac{K(\phi')}{2\left[ \phi' s_s + \left(\frac{\rho_i}{\rho_w} \Delta \phi \right)\right]}\quad &\text{for } \frac{\partial h}{ \partial t}> 0,\\
\kappa_1 = \frac{K(\phi')}{2\phi' (s_s - s_r)}\quad &\text{for } \frac{\partial h}{ \partial t}\leq 0.
    \end{cases}
\end{align}

In this example, the coefficients $\kappa_1$ and $\kappa$ are constant values due to two initially homogeneous media that are inside and outside the aquifer. Similar to classical groundwater theory \citep{Bear_1972} we have used the chain rule to collect head $h$ in the flux term, e.g., $\frac{\partial (h^2/2)}{\partial x} = h\frac{\partial h}{\partial x}$.

We consider the evolution of a finite ``volume'' of liquid water in the aquifer $V(t)$ defined in cartesian (in m$^2$) and cylindrical coordinates (in m$^3$) as 

\begin{align}\label{eq:vol}
    V(t) = \phi' s_s \int_0^{x_{max}(t)} h(x,t)\, \d x \quad \text{and} \quad V(t) =   \phi' s_s 2 \pi \int_0^{r_{max}(t)}r h(r,t)\, \d r, 
\end{align}
respectively, with $x_{max}$ and $r_{max}$ being the maximum horizontal extents of the aquifer (m).  


At late times, the expansion of the firn aquifer becomes independent of the initial conditions and approaches a self-similar solution. As $\kappa_1 \neq \kappa$ because of meltwater freezing and residual trapping, the similarity solution to this problem involves an incomplete similarity, i.e., self-similarity of the second kind (see \cite{Barenblatt1996} for a pedagogical introduction to self-similarity). The self-similar solution of Equations (\ref{eq:governing}-\ref{eq:vol}) can be expressed as

\begin{align}\label{eq:self-similar}
    h = \frac{B^2}{\kappa_1 t^{1-2\beta}} \Phi \left( \zeta, \frac{\kappa}{\kappa_1} \right), \quad x \, \text{ or } \, r = \zeta Bt^\beta
\end{align}
in cartesian or cylindrical coordinates, respectively, where $\zeta$ is the similarity variable and $\beta$ is determined numerically. Here $\Phi$ is the dimensionless function for the height and $B$ is a constant dependent on the initial volume, type of geometry, and the initial horizontal extent $x_{max}$ or $r_{max}$ of the aquifer. In cartesian coordinates $ B = (Q_0\kappa_1)^\beta x_{max,0}^{1-3\beta} $ and in cylindrical coordinates $B = (Q_0\kappa_1)^\beta r_{max,0}^{1-4\beta}$, where the subscript $0$ refers to the values calculated at the initial time $t_0$. The maximum aquifer height scales with time as $h_{max}\sim t^{2\beta-1}$ whereas the maximum horizontal extent or the radius of the aquifer scales with time as $x_{max}$ or $r_{max}\sim t^{\beta}$. The similarity solution \eqref{eq:self-similar} converts the partial differential equation \eqref{eq:governing} and condition \eqref{eq:governing2extra} for aquifer height $h$ to ordinary differential equations for the dimensionless self-similar variable $\Phi$, as summarized in Appendix~\ref{sec:appC_ODEs}. The constant $\beta$ is determined as an eigenvalue to the resulting system of ordinary differential equations when solved simultaneously while satisfying the scaled versions of the boundary conditions of $h=0$ at $x=x_{max}$ or $r=r_{max}$, no-flow at $x=0$ or $r=0$, and the continuity between the two domains at $\partial h / \partial t = 0$. The solutions $\beta$ and $\Phi$ depend on the ratio $\frac{\kappa}{\kappa_1}  = \frac{\phi' (s_s - s_r)}{\phi' s_s + \left(\frac{\rho_i}{\rho_w} \Delta \phi \right)}$ which depends on initial porosity, initial temperature, and residual liquid water trapping. Furthermore, the liquid water volume within the aquifer scales as $V(t) \sim t^{3\beta -1}$ for the cartesian case and $V(t) \sim t^{4 \beta - 1}$ for the cylindrical case. 

The governing equations (Equations \ref{eq:governing}-\ref{eq:vol}) and the resulting ordinary differential equations, given in Appendix~\ref{sec:appC_ODEs}, are identical to the classical problems of finite volume groundwater aquifer expansion with residual trapping for cylindrical coordinates \citep{kochina1983groundwater} and for Cartesian coordinates \citep{huppert1995gravity}. The effect of residual liquid water trapping in terrestrial groundwater models was also accounted for in the definition of $\kappa_1$ in Equation~\eqref{eq:condition-Kochina} in these models \cite[e.g.,][]{kochina1983groundwater,huppert1995gravity}. However, thermodynamics and phase change have never been included in this problem in the context of firn aquifers. Our definition of $\kappa$ includes the porosity reduction when the aquifer expands in the cold firn. The rescaling of the ratio $\kappa/\kappa_1$ thus helps study the aquifers expanding in cold firn while experiencing  liquid water loss due to freezing, closure of pore space, and residual liquid water trapping.

 \subsection{ Comparison against quasi-2D numerical solutions}\label{sec:firnaquiferexpan2D}
We first compare the quasi-two-dimensional solutions from the numerical model against the semi-analytical solution \eqref{eq:self-similar} of the Equations~(\ref{eq:governing}-\ref{eq:vol}, scaled form ~\ref{eq:89_cart}-\ref{eq:91_cart}) in cartesian coordinates. We refer to the dimensions as ``quasi'' because the vertical dimension is represented through vertical averaging. The firn is assumed to be uniform outside the aquifer, with a temperature of $T_0 = -30\,^\circ$C and an initial porosity of $\phi_0=0.7$, which corresponds to $\kappa / \kappa_1=0.925$ and $\Delta \phi = 0.057$ (Figures~\ref{fig2:analyvsnum}\textit{a}, \textit{d}, \textit{e}). We initialize with the semi-analytical solutions in cartesian ($x$) coordinates at time $t=1$ year with an initial maximum horizontal extent of $x_{max}(t=0)=x_{max,0}=2900$ m and $h_{max}(t=0)=200$ m. For the numerical solutions from the vertically integrated firn aquifer model, the horizontal domain $x \in [0,2.5\,x_{max,0}]$ is divided uniformly into 250 grid cells. The boundary conditions are no flow at $x=0$ and a constant aquifer height at the other end, i.e., $h(2.5\,x_{max,0},t)=0$ (Figure~\ref{fig2:analyvsnum}\textit{a}). For validation purposes, we have also included the solutions from a higher-fidelity firn hydrologic model called HydroFirn (for more information, see Appendix~\ref{appA}). HydroFirn solves Equations~(\ref{eq:comp-conservation-final}-\ref{eq:darcy-full}) in two-dimensions and includes heat conduction (contour plots in Figures~\ref{fig2:analyvsnum}\textit{a-i}). The domain for the HydroFirn simulations is $x \in [0,2.5\,x_{max,0}] \times z\in [0,250\text{ m}]$ which is uniformly divided into $80\times40$ grid cells.

The resulting unconfined aquifer expands laterally over time into the cold firn at -30$\, ^\circ$C (Figure~\ref{fig2:analyvsnum}\textit{a}). For this specific firn with calculated $\kappa / \kappa_1=0.925$ based on initial condition, the radial expansion scaling exponent $\beta = 0.328$ is slightly lower than in the case of temperate firn $\beta = 1/3$ in cartesian coordinates. Here liquid water freezing causes a simultaneous reduction in porosity ($\Delta \phi = 0.057$), hydraulic conductivity, and hydraulic head (water level), leading to slower lateral propagation of the aquifer. In a span of 4 years (i.e., $t=5$ years), the aquifer height has reduced to about 120 m, the horizontal extent has increased to 4900 m (Figure~\ref{fig2:analyvsnum}\textit{b}), and the aquifer gradually slows with time as the hydraulic gradients decrease. In a span of about 10 years, the aquifer's height has halved to $\sim$90 m, and its horizontal extent has almost doubled to $\sim$6200 m (Figure~\ref{fig2:analyvsnum}\textit{c}).

The simulation results from the HydroFirn model show comprehensive firn aquifer dynamics and serve as a source of validation (Figures~\ref{fig2:analyvsnum}\textit{a-i}, contour plots). As the aquifer expands, freezing occurs in the 0$\,^\circ$C region that extends into the cold region. Both the semi-analytic and numerical solutions of the vertically integrated model show excellent agreement with the higher-fidelity HydroFirn model. The effect of heat conduction, shown by diffused contours near $z=$200 m in Figures~\ref{fig2:analyvsnum}\textit{d-f}, is far from the top of the aquifer. This is a clear case where heat advection along with the aquifer is much faster than heat conduction. The quantitative estimates of dimensionless aquifer height $h_{max}$, horizontal extent $x_{max}$, and liquid water volume $V$ also show excellent agreement with the semi-analytic solution (Figures~\ref{fig2:analyvsnum}\textit{j-l}). The liquid water volume has decreased by about 4\% due to meltwater freezing (Figure~\ref{fig2:analyvsnum}\textit{l}) with a slightly higher liquid water loss from the HydroFirn model. The higher loss due to freezing caused by heat conduction occurs close to the regions ($x\sim 2500$ m) where the aquifer height does not change significantly ($\partial h/\partial t \approx 0$). Lastly, the HydroFirn model took $\sim$2~hours to simulate upto $t=10$~years, whereas the numerical simulator for the vertically integrated model completed the same simulation in approximately $\sim$6~minutes, both on a single CPU core of an Apple M3 Pro processor with 36~GB RAM. This shows a $\sim$20 times speed-up for this problem when utilizing the vertically-integrated numerical model compared to the higher-fidelity HydroFirn model.


\begin{figure}
    \centering
    \includegraphics[width=0.6\linewidth]{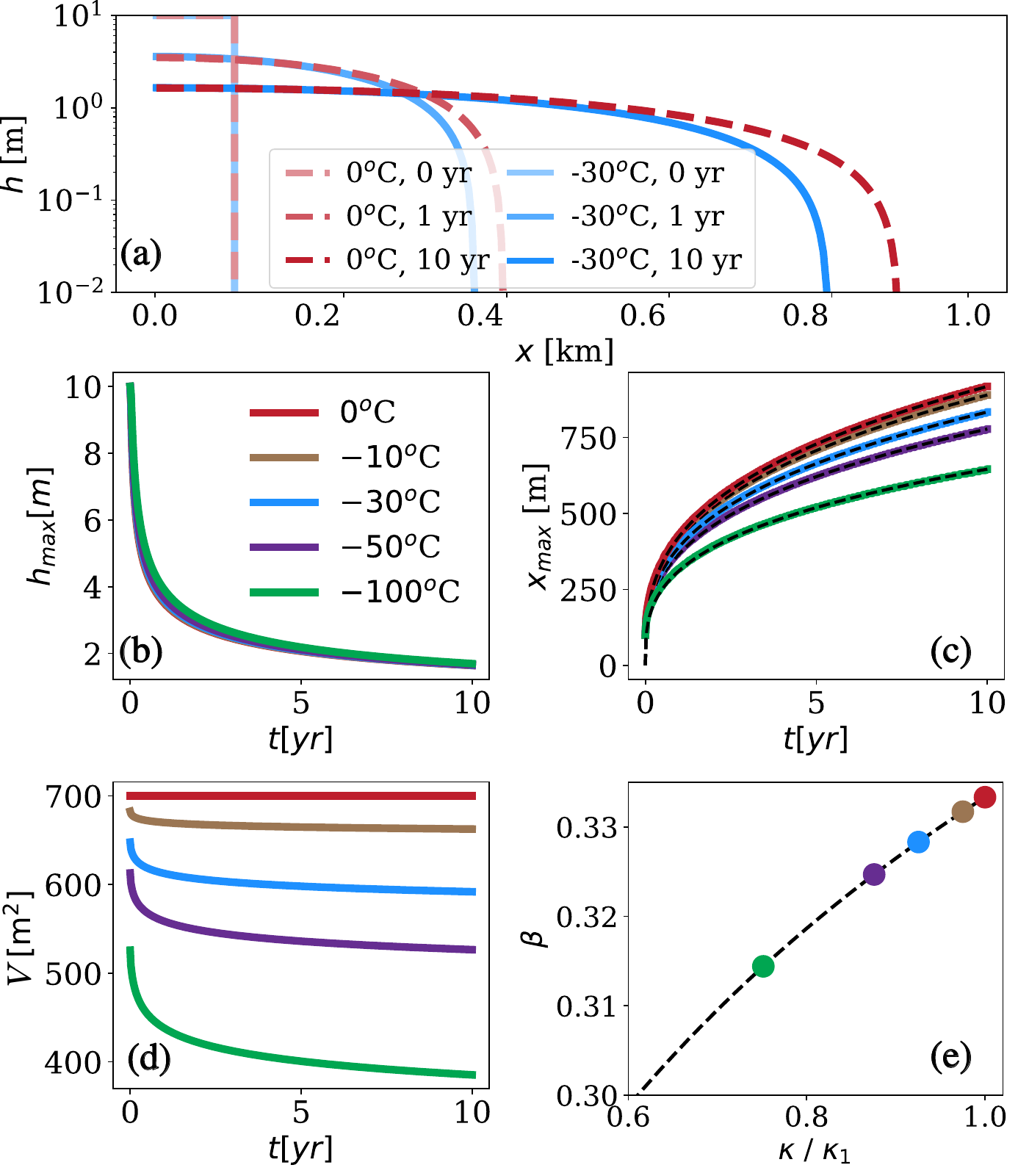}
    \caption{Numerically modeled expansion of an initial, 10 m $\times$ 100 m column of meltwater in cold firn in cartesian coordinates. (a) Evolution of the aquifer height with horizontal distance for the initial firn temperatures $T_0=0\,^\circ$C and $T_0=-30\,^\circ$C. Effect of temperature on the evolution of (b) maximum aquifer height $h_{max}$, (c) maximum horizontal extent of aquifer $x_{max}$ (scalings from the numerical solutions are indicated by markers), (d) liquid water volume within aquifer $V$, and (e) radial expansion scaling exponent $\beta$, i.e., $r_{max}\sim t^\beta$. The dashed black lines in panels \textit{c} and \textit{e} correspond to the theoretical scaling derived for the cartesian case in Section \ref{sec:solutions}. The reduction in porosity corresponding to $\phi_0=0.7$ and $T_0=$0, -10, -30, -50, and -100$^\circ$C is $\Delta \phi =$ 0, 0.018, 0.057, 0.094, and 0.189, respectively. The simulations have same initial total water volume (liquid water + solid ice) in the initial region and have no residual saturation. It is clear that the colder firn slows lateral propagation of aquifers.}
    \label{fig3:effect_of_temp_cartesian}
\end{figure}

Since the numerical simulator is verified and the model is validated, the remainder of this manuscript utilizes numerical results from the vertically integrated model. Next we initialize with a case of a 10 m high by 100 m long column of water that is left to evolve over time (Figure~\ref{fig3:effect_of_temp_cartesian}\textit{a}). This initialization is equivalent to infiltration in a specific region in firn, which is typically much faster than aquifer dynamics, and subsequently allows it to propagate laterally as a firn aquifer. In this case the domain $x\in[0,1000 \text{ m}]$ is divided uniformly into 100 grid cells. The firn outside the aquifer has a porosity of $\phi=0.7$ and is at a uniform temperature of $-30\,^\circ$C ($\Delta \phi = 0.057$). Figure~\ref{fig3:effect_of_temp_cartesian} shows the effect of initial lower temperatures of the firn on the propagation of the aquifer. In this case, colder firn leads to slower propagation of the aquifer and the effect of freezing is more prominent in the regions close to the advancing edge of the front where $\partial h / \partial t>0$ (Figures~\ref{fig3:effect_of_temp_cartesian}\textit{a}, \textit{c}) than in the regions experiencing drainage where $\partial h/\partial t \leq 0$ (Figures~\ref{fig3:effect_of_temp_cartesian}\textit{a}, \textit{b}). Within a year, there is a difference of about 30 m between the horizontal extents of the aquifers expanding in cold and temperate firn due to freezing caused by a thermal deficit ($T_m-T_0$) of 30$\,^\circ$C (Figure~\ref{fig3:effect_of_temp_cartesian}\textit{a}). The theoretical scaling derived for finite volume aquifers in Section~\ref{sec3:const-vol} shows an excellent comparison against the numerical solutions (Figures~\ref{fig3:effect_of_temp_cartesian}\textit{c}, \textit{e}). Also, the initial volumes of liquid water within the aquifer $V(t=0)$ are different for each temperature due to the initial freezing that has caused a reduction of the pore space (Figure~\ref{fig3:effect_of_temp_cartesian}\textit{d}). When firn is colder, the loss of water increases due to freezing while expanding into cold regions.
Furthermore, the horizontal expansion scaling exponent $\beta$ depends on the ratio $\kappa / \kappa_1$ that varies with the firn temperature (Figure~\ref{fig3:effect_of_temp_cartesian}\textit{e}). The parameter $\beta$ goes below its maximum value (1/3 for cartesian geometry) when $\kappa / \kappa_1<1$ in a cold firn. Based on theoretical scaling, it is clear that both the thermal deficit ($T_m - T_0$) and residual trapping ($s_r\neq 0$) will reduce the ratio ${\kappa}/{\kappa_1}$, a factor that also depends on the initial porosity of the firn. This demonstrates that colder firn with residual water trapping leads to the slower propagation of firn aquifers. 

 \subsection{ Comparison against quasi-3D numerical solutions}
\begin{figure}
    \centering
    \includegraphics[width=0.5\linewidth]{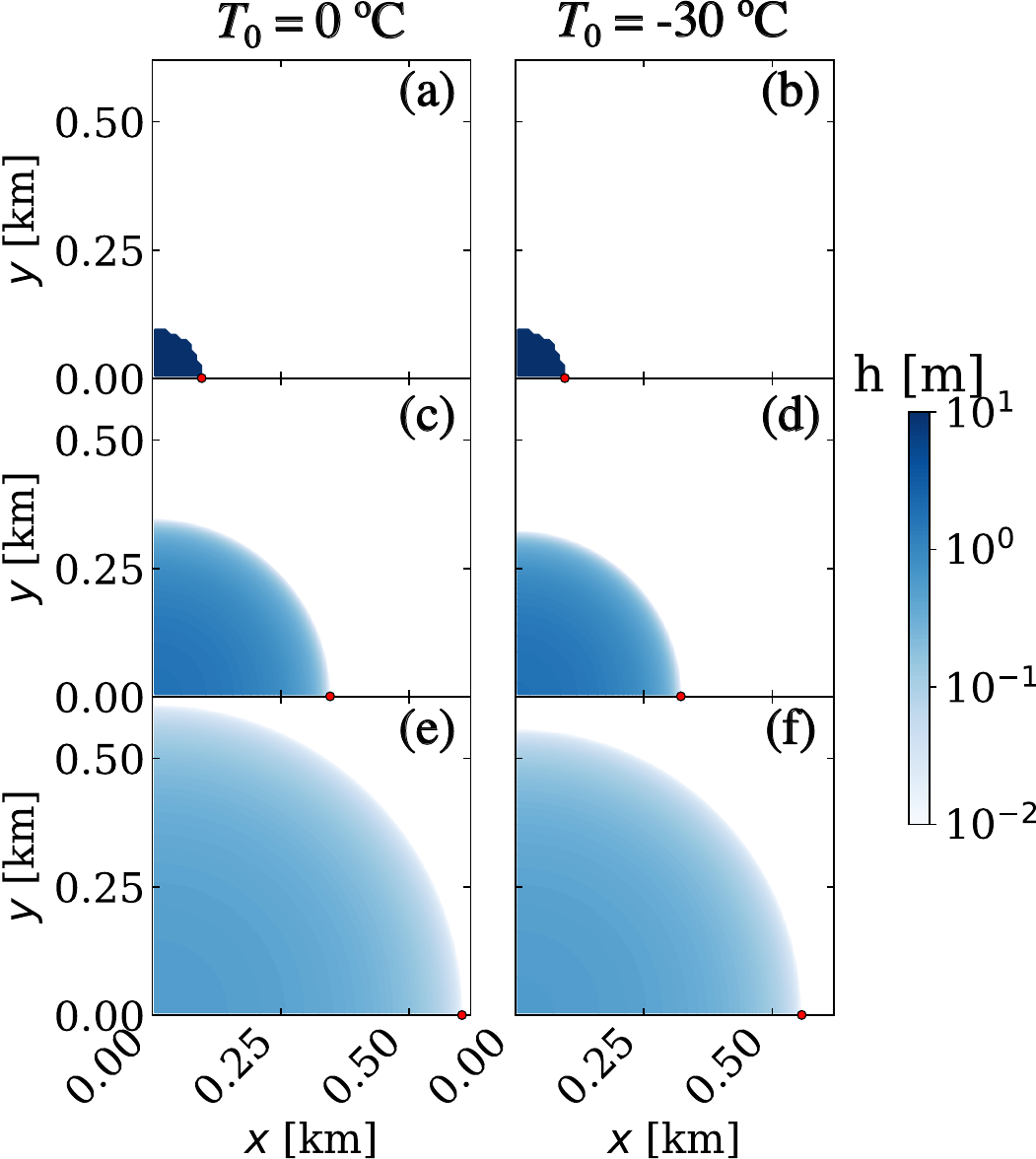}
    \caption{Expansion of an initial, 10 m high column of water with 100 m radius column of meltwater in cold firn in cartesian coordinates. (a) Evolution of the aquifer height with horizontal distance for the initial firn temperatures $T_0=0\,^\circ$C at (a) 0, (c) 1, and (e) 10 years and $T_0=-30\,^\circ$C at (b) 0, (d) 1, and (f) 10 years. Note the log scale for the colorbar in this figure. The aquifer propagates more slowly and attains a smaller spatial extent in cold firn (panels \textit{d}, \textit{f}) compared to the temperate firn case (panels \textit{c}, \textit{e}) at all times after the initial release.
}
    \label{fig4:effect_of_temp_cyl}
\end{figure}

Next we model a quasi-3D firn aquifer in cartesian coordinates ($x,y,z$) and compare the numerical solutions of the vertically integrated model against the semi-analytical solution \eqref{eq:self-similar} of the Equations~(\ref{eq:governing}-\ref{eq:vol}, scaled form ~\ref{eq:89cyl}-\ref{eq:91cyl}) developed for the axisymmetric case (cylindrical $r,z$ coordinates). We initialize the model with a 10~m high column of water with a radius of 100~m, which is then allowed to evolve over time (Figure~\ref{fig4:effect_of_temp_cyl}\textit{a}). This setup is similar to the test shown in Figure~\ref{fig3:effect_of_temp_cartesian}\textit{a}, except that the expansion here is radially outward in two dimensions ($x,y$). In this case the domain $x\in[0,1000 \text{ m}] \times y \in [0,1000 \text{ m}] $ is uniformly divided into $100\times 100$ cells. The firn outside the aquifer has a porosity of $\phi=0.7$ and is at a uniform temperature of $T_0 \leq T_m$. The boundary conditions are no-flow everywhere. 

\begin{figure}
    \centering
    \includegraphics[width=0.6\linewidth]{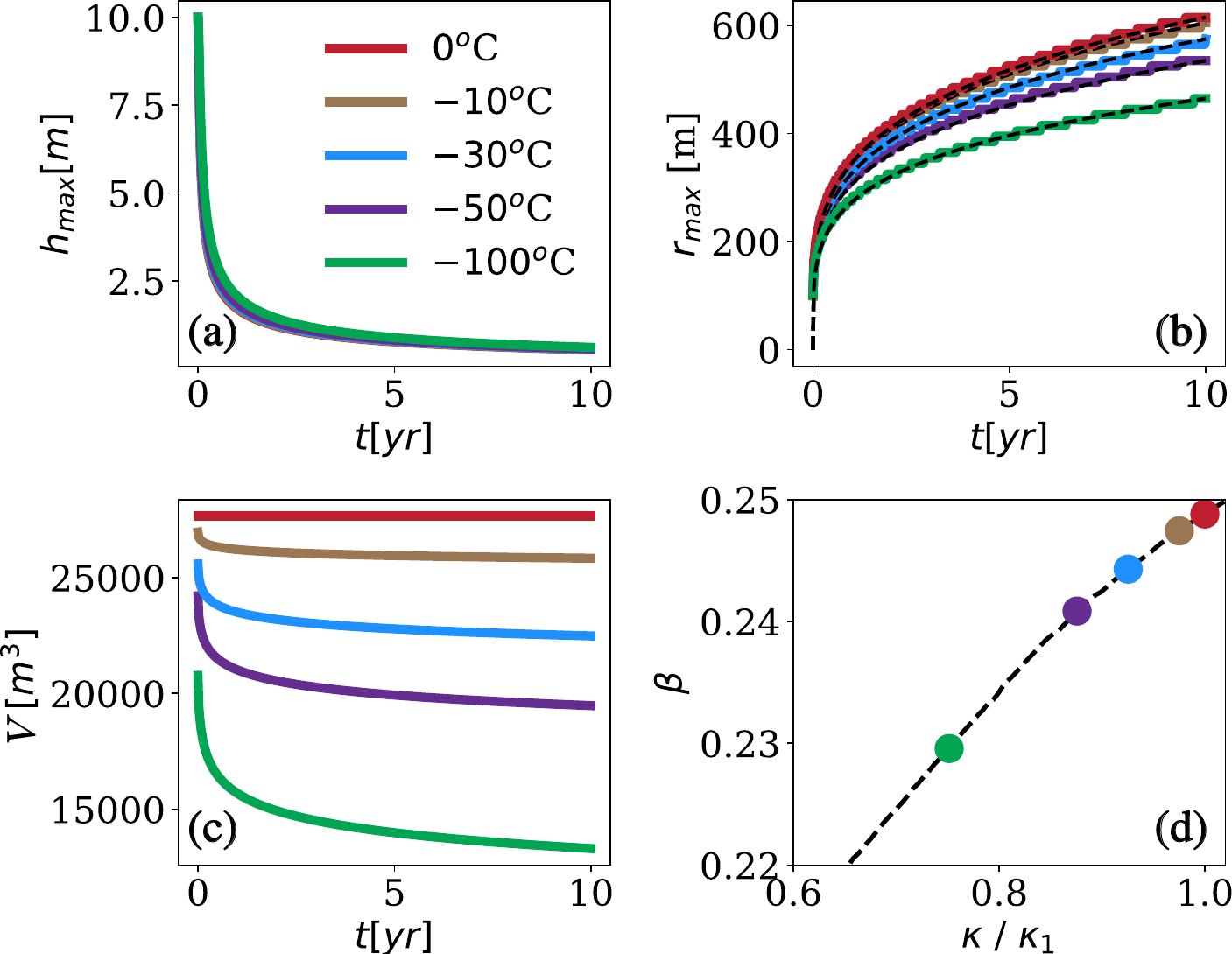}
    \caption{Effect of temperature on the numerically modeled evolution of (a) maximum aquifer height $h_{max}$, (b) maximum horizontal extent of aquifer $r_{max}$, (c) liquid water volume within the aquifer $V$, and (d) radial expansion scaling $\beta$ (the numerically estimated scaling exponents are indicated by markers), corresponding to the quasi-3D simulations reported in Figure~\ref{fig4:effect_of_temp_cyl}. The dashed black lines in panels \textit{b} and \textit{d} correspond to the theoretical scaling developed in cylindrical coordinates in Section \ref{sec:solutions}. The maximum radius of the aquifer $r_{max}$ in panel \textit{b} is evaluated along the $x$-axis, i.e., $y=0$ shown by red dots in Figure~\ref{fig4:effect_of_temp_cyl}. The reduction in porosity corresponding to $\phi_0=0.7$ and $T_0=$0, -10, -30, -50, and -100$^\circ$C is $\Delta \phi =$ 0, 0.018, 0.057, 0.094, and 0.189, respectively. The simulations have the same initial total water volume (liquid water + solid ice) in the initial region and have no residual saturation. It is clear that the colder firn slows unconfined aquifers when compared to the warmer firn.}
    \label{fig4:effect_of_temp_cyl_effects}
\end{figure}

The expansion of the 3D firn aquifer is shown in Figure~\ref{fig4:effect_of_temp_cyl}. We find that at the end of 10 years, the aquifer propagating into cold firn at $T_0=$-30$\,^\circ$C leading to a porosity reduction of $\Delta \phi = 0.057$ (Figures~\ref{fig4:effect_of_temp_cyl}\textit{e}) will be slower than the aquifer expanding in temperate firn (Figures~\ref{fig4:effect_of_temp_cyl}\textit{f}). The finite volume aquifers slow down significantly over time as the horizontal fluxes decrease, diminishing the temperature difference at later stages (Figure~\ref{fig4:effect_of_temp_cyl_effects}\textit{c}). The analysis of firn aquifer dynamics in three dimensions (Figure~\ref{fig4:effect_of_temp_cyl_effects}) at different temperatures demonstrates similar points to those raised previously for the quasi-2D cartesian case (Figure~\ref{fig3:effect_of_temp_cartesian}). The numerical and theoretical solutions in cylindrical coordinates show an excellent comparison.  Furthermore, Figure~\ref{fig4:effect_of_temp_cyl_effects}\textit{d} shows the dependence of the radial expansion scaling exponent $\beta$ on the parameter $\kappa / \kappa_1$, which again depends on the thermal deficit, residual trapping, and initial porosity of firn. The parameter $\beta$ goes below its maximum value (1/4 for cylindrical geometry) when $\kappa / \kappa_1<1$ in cold firn. This again mathematically shows that colder firn, with residual meltwater trapping, leads to slower propagation. In the case of cylindrical geometry, the propagation of firn aquifers is slower compared to the cartesian (quasi-2D) case due to the geometrical nature of aquifer expansion, as was also noted in finite volume aquifer expansion in temperate soils \citep{huppert1995gravity,kochina1983groundwater},

\begin{figure}
    \centering
    \includegraphics[width=0.8\linewidth]{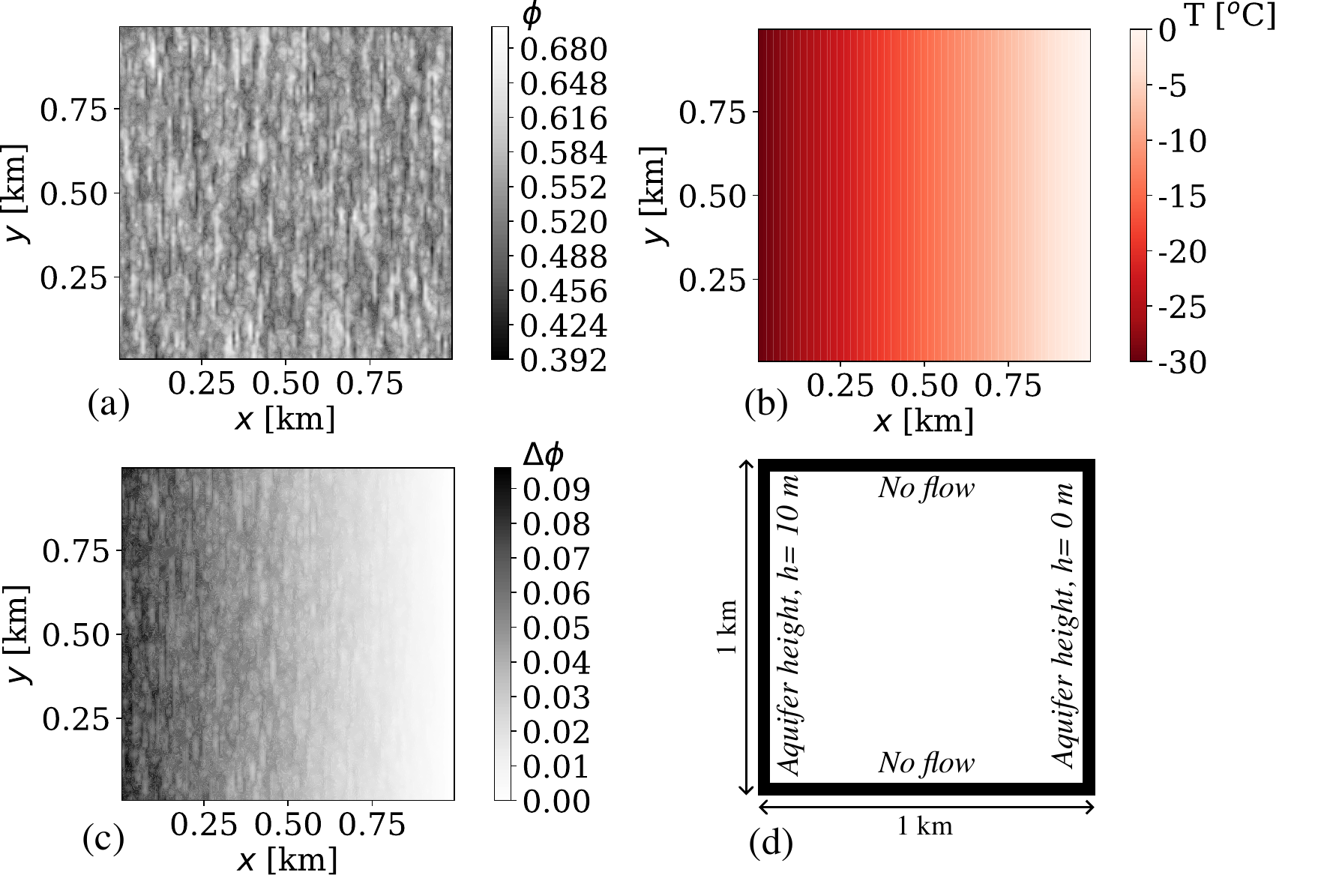}
    \caption{Initial properties of layered heterogeneous firn: (a) porosity field, (b) temperature, and (c) maximum drop in porosity. (d) A schematic of the domain with the boundary conditions. }
    \label{fig6:hetero_init}
\end{figure}

\begin{figure}
    \centering
    \includegraphics[width=0.5\linewidth]{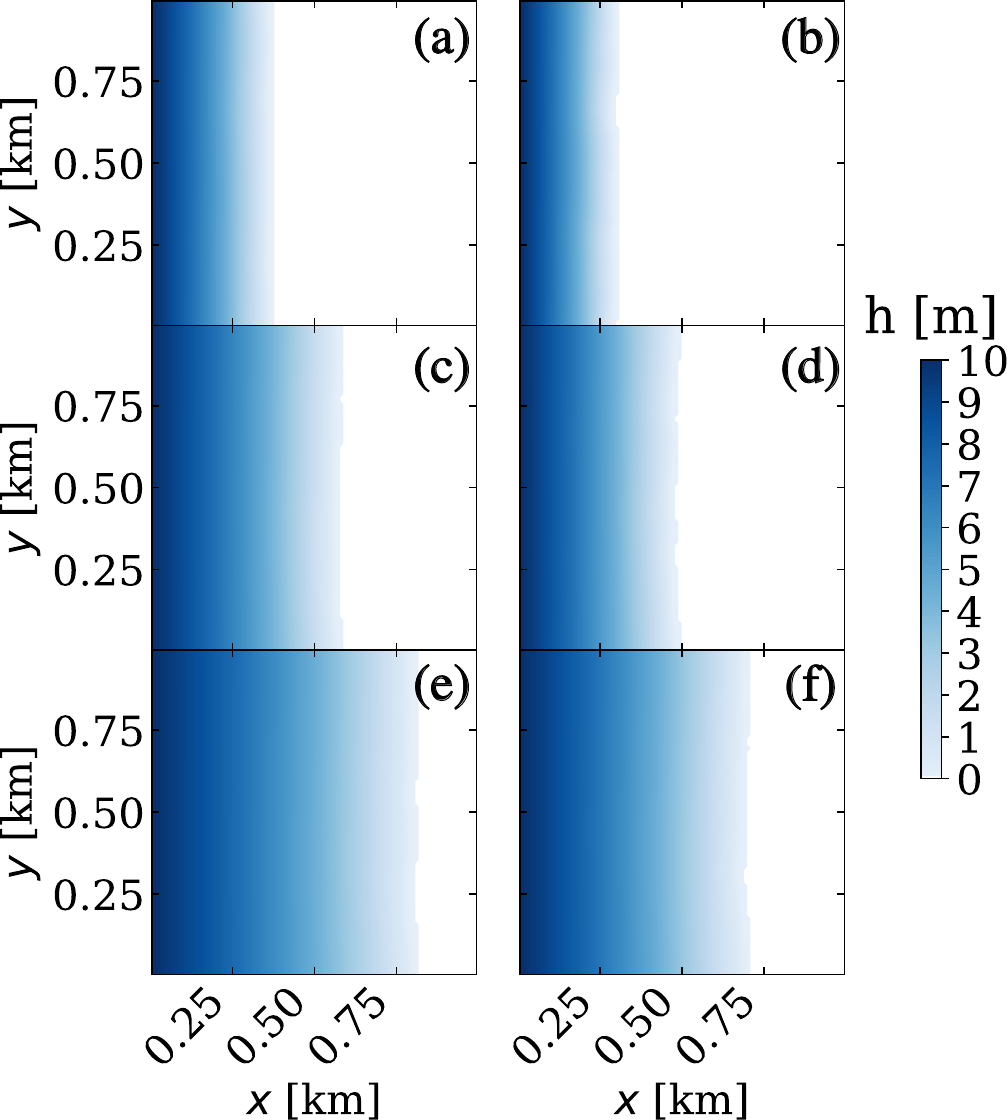}
    \caption{Modeled expansion of aquifer propagation in the temperate firn at 0$\,^\circ$C at $t=$ (a) 1, (c) 2.5, and (e) 5 years and in cold firn with initial temperatures given in Figure \ref{fig6:hetero_init}\textit{b} at (b) 1, (d) 2.5, and (f) 5 years. The level at $h=0$ m is thresholded to white color for improved contrast. Supplementary video S2 shows the corresponding expansion of firn aquifer in temperate and cold firn.}
    \label{fig7:aquiferexpansion_heterogeneous_firn}
\end{figure}

\section{ Aquifer propagation in a heterogeneous, cold firn}\label{sec4:corr-rnd-field}
The final numerical test involves quasi-3D lateral propagation of a firn aquifer in an initially dry and cold heterogeneous firn. To account for natural variability in firn properties, the vertically averaged porosity field is initialized as a correlated random field (Figure~\ref{fig6:hetero_init}\textit{a}). Horizontal correlation scales in firn span from centimeters to hundreds of meters based on stratigraphic observations and may extend to even larger scales due to large-scale accumulation variability \cite[e.g., ][]{laepple2016layering,xu2023polar}. These short and long correlation scales may arise from processes such as layering, temporal variability, and annual snowfall accumulation \citep{xu2023polar}. The anisotropy in porosity, leading to streaks in the $y$ direction, arises from the correlation lengths, chosen to be 10 m and 100 m in the $x$ and $y$ directions, respectively. Furthermore, the amplitude of the logarithmic variation (in powers of ten) is set to 0.1, with a mean of zero. More information on the generation of the correlated random field can be found in \cite{shadab2024hyperbolic}. The domain of dimensions 1000 m $\times$ 1000 m is divided uniformly into 100 $\times$ 100 cells. The boundary conditions on the left and right faces are aquifer heights $h=10$ m and $h =0$ m, respectively, whereas the boundary condition on the rest of the faces is no flow (Figure~\ref{fig6:hetero_init}\textit{d}). There is a linear increase in the initial temperature of the firn from -30$\,^\circ$C to 0$\,^\circ$C from left to right (Figure~\ref{fig6:hetero_init}\textit{b}). 

The thermal deficit ($T_m - T_0(\tilde{\textbf{x}})$, Figure~\ref{fig6:hetero_init}\textit{b}) could lead to a reduction in porosity $\Delta \phi(\tilde{\textbf{x}})$ that is more prominent on the left side of the domain (Figure~\ref{fig6:hetero_init}\textit{c}) due to much colder firn and negligible on the right boundary as the firn temperature approaches $0\,^\circ$C. We compare two cases of aquifer expansion in cold and temperate firn to illustrate the effect of firn temperature on aquifer dynamics. This case is different from the prior cases as the hydraulic head is fixed and the supply of liquid water is infinite for this case. It leads to firn aquifer propagation from left to right (Figure~\ref{fig7:aquiferexpansion_heterogeneous_firn}). The temperature field initially at $T_0(\tilde{\textbf{x}})$ evolves to $T_m$ when the aquifer invades the cold firn, similar to Figures~\ref{fig2:analyvsnum}\textit{d-f}, which is not shown for brevity. Due to heterogeneities, the aquifer moves in a non-planar fashion. It is clear that the cold firn leads to slower propagation of the aquifer (Figure~\ref{fig7:aquiferexpansion_heterogeneous_firn}). In one year, the difference between the leading edge of the aquifers with and without freezing is about 80 m, which grows to about 100 m after 2.5 years (Figure~\ref{fig7:aquiferexpansion_heterogeneous_firn}\textit{f}). This result shows the importance of considering thermal deficit and freezing caused by heat advection in firn aquifer models.

\section{ Discussion and limitations}\label{sec5:discussion-limitations}
The theoretical framework and the associated numerical model  developed in this paper provide a foundation for understanding the interplay between aquifer dynamics, heat transport, and phase change in cold firn. In particular, for problems where the firn is initially warmed to the melting temperature when invaded, the resulting change in porosity $(\Delta \phi)$ and reduction in liquid water volume can be leveraged as a condition to solve a range of firn aquifer problems, such as late-stage drainage from firn aquifers or the expansion of firn aquifers due to constant flux. Furthermore, for a steady state the same equations could be used with a reduced porosity such as a steady aquifer in firn. A compiled set of such problems in temperate soil is presented in \cite{shadab2024hyperbolic} that offers a suite of simplified yet physically consistent formulations that can be employed as benchmark solutions. These limiting-case scenarios are especially useful for verifying and validating higher-fidelity firn hydrologic models such as \cite{miller2022empirical}, where coupled thermodynamics, multiphase flow, and evolving microstructure interact in complex ways. Further, in addition to being implemented in prior models for temperate firn, the present model extends the framework to problems where firn aquifers invade cold regions, a process that has not been studied previously and will be increasingly important due to warming causing the expansion of firn aquifers, as suggested by studies such as \cite{horlings2022expansion,brils2024climatic}.

 Compared to the theoretical solutions, the results from the numerical models account for spatial heterogeneity in variables such as temperature, porosity, and hydraulic conductivity. Allowing each computational cell to possess distinct thermophysical properties captures the natural variability present in firn columns, where layering and firn metamorphism modulate energy and water mass transfer. Heterogeneity in firn properties, such as ice layers, can form and/or affect the propagation of firn aquifers.
Thin, discontinuous ice layers may not substantially impede lateral spreading, allowing the aquifer to remain largely unconfined, as observed in \cite{killingbeck2020integrated}. In contrast, thicker and more continuous ice slabs can compartmentalize the pore space, leading to the formation of confined aquifers and localized pressurization. 

For the development of the vertically integrated model, the sharp-interface approximation is crucial, provided that the Peclet numbers associated with liquid water and heat transport are considered large. The Peclet number is defined as the ratio of the advective transport to the diffusive transport of a quantity. In the unsaturated region above the water table, diffusion due to capillary forces can lead to the suction of water out of the water table, forming unsaturated regions with saturation $s_s<s_w<s_r$. For large scales and regions where the Peclet number for hydrology $Pe_{w}$ is high, i.e., when gravitational forces dominate over capillary forces, the sharp interface approximation is justified, leading to no unsaturated regions. As a result, the saturation can only take two values, i.e., $s_w\in \{s_r,s_s \}$. This is a good approximation for large spatial scales and firn having a larger grain size. A more detailed discussion on variably saturated flows with sharp interfaces can be found in \cite{shadab2022analysis,shadab2024hyperbolic} for soil and \cite{shadab2024mechanism,shadab2025unified} for glacial firn. 

Next, we discuss the sharp-interface assumption and its limitations for heat transport. The thermal diffusivity of ice $\alpha \approx 10^{-6}$ m$^2$/s at 0$\,^\circ$C \citep{james1968thermal}. For porous firn, a prefactor of 0.25-1 could be applied to the thermal diffusivity to account for the thermal diffusivity of cold firn for a porosity between 0 and 0.7 (e.g., the factor being $(1-\phi)^{l-1}$; see \cite{shadab2024mechanism} with exponent $l=1.885$ taken from \cite{Yen1981new}). In the vertical direction, the heat conduction length scales from the scaling $\sqrt{\alpha t}$ are on the order of one meter at yearly timescales. If the length scales for heat advection along with the aquifer motion are on the order of ten meters or more in a year, conduction could be considered small at the annual time scales, as observed in Figures~\ref{fig2:analyvsnum}\textit{g}-\textit{i}. But if the vertical motion of the aquifer is on the order of one meter per year or less, while the time of consideration is on the order of years, then heat conduction would be prominent. In the case where heat conduction is dominant, the loss of liquid water due to conduction loss could be accounted for as a sink term in the governing equation (\ref{eq:governing2}). This is an excellent direction for future work.

Latent heat provides dominant thermal buffering for aquifer thermodynamics. Because water's latent heat of fusion far exceeds firn's sensible heat capacity, small amounts of liquid water freezing warm the surrounding firn and maintain near-isothermal conditions at the melting point. This self-limiting mechanism for conductive heat loss keeps deep firn aquifers thermally stable despite seasonal surface forcing. Field observations confirm this behavior: deep firn aquifers show minimal seasonal temperature variation and remain near the melting point year-round \citep{koenig2014initial,miller2018direct,miller2020hydrology,miller2023hydrologic,van2025long}. For the firn where the aquifer reaches close to the surface, climate effects on the firn aquifer dynamics are much more prominent and need to be accounted for, as pointed out in \cite{van2025long}. This new vertically integrated model can easily be coupled with a 1D firn model for localized heat conduction, as was done in \cite{van2025long}. A source term in the model equation \eqref{eq:governing2} can account for aquifer recharge but will lead to more intricate thermodynamics impacting the maximum aquifer thickness $h_{max}$, defined in Equation \eqref{eq:hmax}.

Importantly, these insights demonstrate the efficiency and utility of vertically integrated models in capturing the bulk behavior of firn aquifers while maintaining computational tractability. By integrating the governing equations over depth, vertically integrated models naturally incorporate effective source terms derived from surface boundary conditions in the full thermodynamic-hydrologic model, linking recharge, drainage, and freezing processes in a reduced-dimensional framework. This approach preserves the dominant physics, particularly mass and energy conservation, while substantially reducing the degrees of freedom relative to fully three-dimensional multiphase flow models. As a result, vertically integrated formulations for firn aquifers offer a powerful balance between physical fidelity and computational efficiency, making them ideally suited for long-term simulations, sensitivity analysis, and benchmarking exercises. For example, the vertically integrated numerical model developed in the present work is about 20 times faster than a higher-fidelity firn hydrologic model (HydroFirn) for the quasi-2D firn aquifer expansion problem studied in Section~\ref{sec:firnaquiferexpan2D}. These firn aquifer models can also serve as foundational components for hybrid models that couple surface energy balance, hydrologic routing, and thermodynamic evolution at larger spatial and temporal scales.

Overall, the results highlight the importance of combining analytical limits, vertically integrated formulations, and numerical experiments to unravel the coupled processes of heat transport, phase change, and evolving porosity in firn hydrology. The simplified problems outlined here, when used as benchmark tests, can help evaluate the fidelity of multiphase flow models under well-defined physical conditions, guiding the development of more robust and predictive firn hydrology simulators.

\section{ Conclusions}\label{sec6:conclusions}
We developed a vertically integrated model framework and an associated numerical model that couples hydrology, heat transport, and phase change for firn aquifers. The framework unifies temperate and cold aquifer systems by explicitly accounting for meltwater freezing caused by heat advection and residual water trapping. Analytical and numerical solutions demonstrate that colder initial temperatures causing meltwater freezing and residual trapping both slow lateral aquifer expansion through enhanced liquid water loss. The solutions compare well against each other and against a higher-fidelity firn hydrologic model. The model framework represents lateral groundwater-style flow with freezing in firn, linking meltwater dynamics, porosity evolution, and heat transport. By integrating over depth, it captures the lateral flow of meltwater and freezing in a reduced-dimensional form that preserves mass and energy conservation while remaining computationally efficient. These results establish a foundation for benchmarking firn hydrologic simulators and investigating how aquifer formation, ice-layer structure, and phase change in firn govern meltwater dynamics and storage. The framework thus offers new insights into the evolution of firn aquifers in and beyond percolation zones and their role in modulating surface mass loss and sea-level rise.

\section*{ Supplementary material}
Supplementary video S1 shows the modeled time evolution of the firn aquifer given by the higher-fidelity HydroFirn simulator with heat conduction (contour plots or solid blue lines) and semi-analytical solutions (red dashed lines) corresponding to Figure~\ref{fig2:analyvsnum}. Furthermore, Supplementary video S2 shows the numerical solutions for the expansion of the aquifer in temperate and cold firn from the vertically integrated model corresponding to Figure~\ref{fig7:aquiferexpansion_heterogeneous_firn}.
\section*{Code availability}
All codes used to generate the analytical and numerical results presented in this manuscript are publicly available on GitHub: \url{https://github.com/mashadab/ColdFirnAquifer3D} \citep{shadab2025coldfirnaq3D}.

\section*{ Funding} 
M.A.S. was supported through Princeton University's Future Faculty in Physical Sciences Postdoctoral Fellowship.

\appendix
\section{ Higher-fidelity firn hydrologic model}\label{appA}

Firn is considered a three phase system comprising porous ice ($i$) with liquid water ($w$) and non-reactive gas ($g$) in the pore spaces. We assume that the gas phase does not contribute to the fluid mechanics or thermodynamics of the system due to its very low density, viscosity, and thermal conductivity. The conserved variables are the total water composition (kg/m$^3$), $C$, defined as the total mass of the water in liquid and solid form per unit representative elemental volume (REV) and the enthalpy of the system (J/m$^3$), $H$ per unit REV due to the phase change involved \citep{shadab2024mechanism,shadab2025unified} defined as

\begin{align} \label{eq:C-working-def}
   C &:= \rho_i \phi_i + \rho_w \phi_w,\\
    H &:= \begin{cases}\rho_i c_{p,i} \phi_i (T-T_m), & {T < T_m} \quad (\textrm{or } H \leq 0) \\  \rho_w \phi_w L,  &{T= T_m} \quad (\textrm{or } 0 < H < CL) \\ \rho_w \phi_w \left(c_{p,w} (T-T_m) + L \right), &{T> T_m} \quad (\textrm{or } H \geq C L) \end{cases}
    ,\label{eq:enthalpy-formulation}
\end{align}
where $\rho_\alpha$ is the density (kg/m$^3$), $\phi_\alpha$ refers to the volume fraction of phase $\alpha \in \{w,i,g \}$,  $T$ is the temperature of the firn (K), and $T_m$ is the melting temperature. Also, $c_{p,\alpha}$ is the specific heat capacity at constant pressure (J/kg$\cdot$K) for phase $\alpha$, $T_m$ is the melting temperature (K) and $L$ is the latent heat of fusion of water (J/kg).
The governing equations correspond to the conservation of these variables as
\begin{align}
    \frac{\partial C}{\partial t} + \nabla \cdot (\textbf{q} \rho_w) &= 0, \label{eq:comp-conservation-final}\\
    \frac{\partial H}{\partial t} + \nabla \cdot (\textbf{q} \rho_w \left(c_{p,w} (T-T_m) + L \right) - \overline{\kappa} \nabla T) &= 0, \label{eq:enthalpy-conservationfull}
\end{align}

where $\textbf{q}$ is the volumetric flux of water phase (m$^3$/m$^2\cdot$s) relative to ice phase. Here $\overline{\kappa}$ is the effective thermal conductivity of the firn.

The volumetric flux of water relative to ice, $\textbf{q}$, can be expressed using Darcy's law,

\begin{align} \label{eq:darcy-full}
    \textbf{q} = - K(\phi,s_w) \nabla h
\end{align}
where $h$ is the hydraulic head (m) and $K(\phi,s_w)$ is the unsaturated hydraulic conductivity of the firn, which is a function of porosity and saturation; however, when the medium is completely saturated ($s_w=s_s$), we refer to it as saturated hydraulic conductivity $K(\phi,s_w=s_s)\equiv K$. The model assumes gravity dominated flow in unsaturated regions, such that the capillary pressure is negligible at the problem length scales. As such, the hydraulic head is assumed to be $h=z$ in unsaturated cells. The numerical model to solve Equations (\ref{eq:comp-conservation-final}-\ref{eq:darcy-full}) using the discrete operator toolbox framework is referred to as the HydroFirn model in this manuscript. See \cite{shadab2024hyperbolic} for a pedagogical introduction to the discrete operator toolbox.

\section{ Derivation of the vertically integrated model for Firn Aquifers from enthalpy balance (\protect\ref{eq:enthalpy-conservationfull})}\label{appB_enthalpy}
 Here we restrict the analysis to temperatures below or at melting temperatures. Due to the sharp interface assumption, the medium has only one mobile phase at any instant, either air or liquid water, but never both; thus, only either the conservation of enthalpy or total water mass (composition) may be enough to represent the system. Below we utilize enthalpy equation \eqref{eq:enthalpy-conservationfull} to derive the vertically integrated model but Appendix~\ref{appsec:composition} uses the conservation of composition (total water mass) for the derivation, both yielding identical equations~(\ref{eq:governing2}-\ref{eq:governing2extra}). 

This leads to the conserved variable being the total enthalpy of the system $H$ (in J/m$^3$) defined as $H:=\rho_i c_{p,i} (T-T_m) \phi_i + \rho_w \phi_w L$ and the governing equation \eqref{eq:enthalpy-conservationfull} simplifies to
\begin{align}
        \frac{\partial H}{\partial t} + \nabla \cdot (\rho_w L \textbf{q} - \overline{\kappa} \nabla T) &= 0, \label{eq:enthalpy-conservation}
\end{align}

 where $\overline{\kappa}$ is the thermal conductivity of the firn (W/m$\cdot$K) and as $\textbf{q}=0$ for $T<T_m$ due to absence of liquid water. Plugging the definition of $H$ and dividing both sides by $\rho_w L$ gives

\begin{align}
        \frac{\partial \left( -\Delta \phi(T) \rho_i/\rho_w+ \phi_w \right)}{\partial t} + \nabla \cdot \left(\textbf{q}  -  \frac{ \overline{\kappa}}{\rho_w L }  \nabla T\right) &= 0, \label{eq:enthalpy-conservation3}
\end{align}

where the reduction in porosity due to freezing is defined as $\Delta \phi(T):=  \frac{c_{p,i} (T_m-T) \phi_i}{L} $. Integrating both vertically from $0$ to $h(\tilde{\textbf{x}},t)$ assuming a negligible change in the vertical direction compared to lateral variations gives

\begin{align}
        \int_0^{h(\tilde{\textbf{x}},t)} \left[\frac{\partial \left( -\Delta \phi(T)\rho_i/\rho_w + \phi_w \right)}{\partial t} + \nabla \cdot \left(\textbf{q}  -  \frac{ \overline{\kappa}}{\rho_w L }  \nabla T\right) \right] \, \d z &= 0
\end{align}
where $\tilde{\textbf{x}}$ are the lateral directions irrespective of the coordinate system.
We assume that the thermal Peclet number, defined as the ratio of heat transport via advection along with the aquifer to heat conduction via temperature gradients, is very high such that the conduction term drops. Assuming no flow in the vertical direction $\nabla \cdot \textbf{q}$ becomes $\tilde{\nabla}\cdot \textbf{q}$ where $\tilde{\nabla}$ is the divergence only in the lateral direction
\begin{align}
        \int_0^{h(\tilde{\textbf{x}},t)} \left[\frac{\partial \left( -\Delta \phi(T)\rho_i/\rho_w + \phi_w \right)}{\partial t} + \tilde{\nabla} \cdot \textbf{q}\right]  \, \d z &= 0,
\end{align}
which is identical to the mass balance of the water phase. Now, for the firn aquifer, we use Darcy's law for transport, given as $\textbf{q}=-K\tilde{\nabla} h$, where $K$ is the saturated hydraulic conductivity in the firn aquifer, and $h$ is the hydraulic head, equivalent to the height of the unconfined aquifer. Here the saturated hydraulic conductivity is evaluated at the porosity $\phi'$. Plugging it in the equation gives

\begin{align}
        \int_0^{h(\tilde{\textbf{x}},t)} \left[\frac{\partial \left( -\Delta \phi(T)\rho_i/\rho_w + \phi_w \right)}{\partial t}  + \tilde{\nabla} \cdot \left(-K\tilde{\nabla} h  \right) \right] \, \d z &= 0.
\end{align}

By integrating both sides, we get 

\begin{align}
        \int_0^{h(\tilde{\textbf{x}},t)} \frac{\partial \left( -\Delta \phi(T)\rho_i/\rho_w + \phi_w \right)}{\partial t} \,\d z+  \int_0^{h(\tilde{\textbf{x}},t)} \tilde{\nabla} \cdot \left(-K\tilde{\nabla} h  \right) \d z &= 0
\end{align}

Considering the flux  term is independent of $z$, we can write

\begin{align}
        \int_0^{h(\tilde{\textbf{x}},t)} \frac{\partial \left( -\Delta \phi(T)\rho_i/\rho_w + \phi_w \right)}{\partial t} \, \d z+ \tilde{\nabla} \cdot \left(-Kh\tilde{\nabla} h  \right) &= 0
\end{align}

Lastly, we will use the Leibniz integral rule to evaluate the time derivative term,

\begin{align}\label{eq:vertically-integrated}
\frac{\partial }{\partial t} \left( \int_0^{h(\tilde{\textbf{x}},t)} \left( -\Delta \phi(T)\rho_i/\rho_w + \phi_w \right) \d z \right) - \left( -\Delta \phi(T)\rho_i/\rho_w + \phi_w \right)_{z=h} \cdot \frac{\partial h}{\partial t} +  \tilde{\nabla} \cdot \left(-Kh \tilde{\nabla} h  \right) = 0
\end{align}

Although discontinuous, we assume the values at the boundary from outside the water table, i.e., $\left( -\Delta \phi_i(T)\rho_i/\rho_w  + \phi_w \right)|_{z=h^+}$. This means $\phi_w|_{z=h^+}$ is 0 when we invade the cold region (Region I with $\partial h/\partial t >0$ and $h(x,t)=h_{max}[\tilde{\textbf{x}},t]$) and $\phi_w|_{z=h^+}$ is equal to $\phi' s_{r}$ for the region experiencing drainage or invading a previously wetted and warmed pore space (Region II and III). Here $\phi'$ is the refrozen space porosity defined as $\phi'=\phi_0 - \Delta \phi(T_0)$ and $s_r$ is the residual saturation. Similarly, for the pore available to freeze $\Delta \phi(T)|_{z=h^+} =\Delta \phi(T_0)$ when we invade a new region (Region I, $\partial h/\partial t >0$ and $h(x,t)=h_{max}[\tilde{\textbf{x}},t]$) otherwise $\Delta \phi(T)|_{z=h^+} =\Delta \phi(T_m)=0$ since the porous medium has already been warmed.

For the accumulation term (first term in Equation \eqref{eq:vertically-integrated}), we use values from inside the aquifer, i.e., $\Delta \phi(T_m) = 0$ and $\phi_w = \phi' s_s$. This greatly simplifies our problem. We remove the tilde in the following analysis for brevity.

\textit{I. When draining from a region (Region II invades Region III; Figure~\ref{fig1:sample-aquifer}\textit{d})}, $\partial h/\partial t \leq 0$:
In this case, the Equation~\eqref{eq:vertically-integrated} simplifies to
\begin{align}
\left(  \phi' s_s \right)\frac{\partial h}{\partial t}  - \left( -\Delta \phi(T_m)\rho_i/\rho_w   +  \phi' \, s_r\right)_{z=h^+} \, \frac{\partial h}{\partial t} +  \nabla \cdot \left( - K h \nabla h \right)  &= 0 \nonumber, \\
\Rightarrow\phi' \, \left(s_s- s_r\right) \, \frac{\partial h}{\partial t} +  \nabla \cdot \left( - K h \nabla h \right)  &= 0.
\end{align}

\textit{II. When invading a new, cold region (Region III invades Region I; Figure~\ref{fig1:sample-aquifer}\textit{e})}, $\partial h/\partial t > 0$ and $h(\textbf{x})=h_{max}[\tilde{\textbf{x}},t]$: In a similar fashion, we obtain
\begin{align}
\phi'  s_s \, \frac{\partial h}{\partial t} + \Delta \phi(T_0)\frac{\rho_i}{\rho_w}  \frac{\partial h}{\partial t} +  \nabla \cdot \left( - K h \nabla h \right)  &= 0 .
\end{align}

\textit{III. When invading a previously invaded region (Region III invades Region II; Figure~\ref{fig1:sample-aquifer}\textit{f})}, $\partial h/\partial t > 0$ and $h(\textbf{x})<h_{max}[\tilde{\textbf{x}},t]$: In this case, we obtain more water from the residual saturation available in the previously invaded region (Region II) leading to
\begin{align}
\phi' \, \left(s_s- s_r\right) \, \frac{\partial h}{\partial t} +  \nabla \cdot \left( - K h \nabla h \right)  &= 0.
\end{align}

\section{ Derivation of the vertically integrated model for Firn Aquifers from Total water mass (composition) balance (\protect\ref{eq:comp-conservation-final})}\label{appsec:composition}

In this case, the conserved variable is the total water mass per unit REV defined as composition $C$ (in kg/m$^3$) defined as $C:=\rho_i \phi_i + \rho_w \phi_w$ and the governing equation \eqref{eq:comp-conservation-final} simplifies to
\begin{align}
    \frac{\partial \left( \rho_i \phi_i + \rho_w \phi_w \right)}{\partial t} + \nabla \cdot (\rho_w \textbf{q}) = 0.
\end{align}

Integrating both vertically from $0$ to $h(\tilde{\textbf{x}},t)$ assuming a negligible change in the vertical direction compared to lateral variations gives

\begin{align}
   \int_0^{h(\tilde{\textbf{x}},t)} \frac{\partial \left( \rho_i  \phi_i  + \rho_w \phi_w \right)}{\partial t} \d z +  \int_0^{h(\tilde{\textbf{x}},t)}  \tilde{\nabla} \cdot (\rho_w \textbf{q}) \, \d z = 0.
\end{align}

where $\tilde{\textbf{x}}$ are the lateral directions, irrespective of the coordinate system. Using the Leibniz integral theorem for the time derivative and the definition of the volumetric flux of liquid water \textbf{q}, we can write

\begin{align}\label{eq:vertically-integrated-comp}
\frac{\partial }{\partial t} \left( \int_0^{h(\tilde{\textbf{x}},t)} \left( \rho_i  \phi_i  + \rho_w \phi_w \right) \partial z \right) - \left( \rho_i  \phi_i  + \rho_w \phi_w \right)_{z=h} \cdot \frac{\partial h}{\partial t} +  \int_0^{h(\tilde{\textbf{x}},t)}  \tilde{\nabla} \cdot \left( - \rho_w K \nabla h \right) \, \d z = 0.
\end{align}

Although discontinuous, we assume the values at the boundary from outside the water table, i.e., $\left( \rho_i  \phi_i  + \rho_w \phi_w \right)_{z=h^+}$. This means $\phi_w$ is 0 when we invade a new, cold region and $\phi_w$ is equal to $\phi' s_{r}$ for the region applies otherwise, i.e., when the aquifer drains or invades a previously warmed region. Here $\phi'$ is the refrozen space porosity, and $s_r$ is the residual saturation. Similarly, for the boundary of the ice volume fraction $\phi_i =(1-\phi_0)$, when the aquifer invades a new, cold region, whereas $\phi_i =(1-\phi')$ applies otherwise. For the accumulation term (the first term in Equation \eqref{eq:vertically-integrated-comp}), we use values from inside the aquifer, i.e., $\phi_i = (1-\phi')$ and $\phi_w = \phi' s_s$. This greatly simplifies our problem. We remove the tilde in the following analysis for brevity.

\textit{I. When draining from a region (Region II invades Region III; Figure~\ref{fig1:sample-aquifer}\textit{d})}, $\partial h/\partial t \leq 0$:
In this case, the Equation~\eqref{eq:vertically-integrated-comp} simplifies to
\begin{align}
\left( \rho_i  (1-\phi')  + \rho_w \phi' s_s \right)\frac{\partial h}{\partial t}  - \left( \rho_i  (1-\phi')  + \rho_w \phi' \, s_r\right)_{z=h^+} \, \frac{\partial h}{\partial t} +  \nabla \cdot \left( - \rho_wK h \nabla h \right)  &= 0 \nonumber \\
 \Rightarrow \phi' \, \left(s_s- s_r\right) \, \frac{\partial h}{\partial t} +  \nabla \cdot \left( -  K h \nabla h \right)  &= 0.
\end{align}

\textit{II. When invading a new, cold region (Region III invades Region I; Figure~\ref{fig1:sample-aquifer}\textit{e})}, $\partial h/\partial t > 0$ and $h(\textbf{x})=h_{max}[\tilde{\textbf{x}},t]$: In a similar fashion, we obtain
\begin{align}
\left( \rho_i  (1-\phi')  + \rho_w \phi' s_s \right)\frac{\partial h}{\partial t}  - \left( \rho_i  (1-\phi_0)  + \rho_w \phi' \, 0\right)_{z=h^+} \, \frac{\partial h}{\partial t} +  \nabla \cdot \left( - \rho_w K h \nabla h \right)  &= 0 \nonumber\\
 \Rightarrow   \Delta \phi \frac{\rho_i}{\rho_w}  \frac{\partial h}{\partial t}  + \phi'  s_s \, \frac{\partial h}{\partial t} +  \nabla \cdot \left( - K h \nabla h \right)  &= 0 \quad \textrm{as $\phi_0 - \phi' = \Delta \phi$}.
\end{align}

\textit{III. When invading a previously invaded region (Region III invades Region II; Figure~\ref{fig1:sample-aquifer}\textit{f})}, $\partial h/\partial t > 0$ and $h(\textbf{x})<h_{max}[\tilde{\textbf{x}},t]$: In this case, we obtain more water from the residual saturation available in the previously invaded region (Region II) leading to
\begin{align}
\phi' \, \left(s_s- s_r\right) \, \frac{\partial h}{\partial t} +  \nabla \cdot \left( - K h \nabla h \right)  &= 0.
\end{align}

\section{ Resulting ordinary differential equations for the expansion of finite volume aquifer in cold firn}\label{sec:appC_ODEs}

\subsection{ Cartesian coordinates}
In cartesian coordinates, we implement the scaling \eqref{eq:self-similar} in the governing equations (\ref{eq:governing}-\ref{eq:condition-Kochina}) and utilize the constraint on finite volume \eqref{eq:vol} and boundary conditions at $x=0$ and $x=x_{max}$ as defined in Section \ref{sec:solutions}. The resulting ODE system for the scaled height $\Phi$ and the eigenvalue $\beta$ comes out to be

\begin{align}\label{eq:89_cart}
   \frac{\d^2 \Phi^2}{\d \zeta^2} + \frac{\kappa_1}{\kappa} \left[\left(1 - 2 \beta \right) \Phi + \beta \zeta \frac{\d \Phi }{\d \zeta} \right] &= 0, \quad \zeta > \zeta_0 \text{ or }\left(1 - 2 \beta \right) \Phi + \beta \zeta \frac{\d \Phi }{\d \zeta}<0\quad \text{and}\\
   \frac{\d^2 \Phi^2}{\d \zeta^2} +\left(1 - 2 \beta \right) \Phi + \beta \zeta \frac{\d \Phi }{\d \zeta}  &= 0, \quad \zeta \leq \zeta_0 \text{ or }  \left(1 - 2 \beta \right) \Phi + \beta \zeta \frac{\d \Phi }{\d \zeta} \geq  0 . \\
\text{subject to boundary conditions}& \hfill \nonumber\\
    \frac{\d \Phi(\zeta=0)}{\d \zeta} = 0, \quad  &  \frac{\d \Phi(\zeta=1)}{\d \zeta}  =- \frac{\kappa_1}{2\kappa} \beta \quad \text{and}\quad \Phi(\zeta=1) = 0.\label{eq:91_cart}
\end{align}
where the dimensionless location $\zeta = 0$ is the axis of the aquifer where its height is maximum, corresponding to the horizontal location $x=0$, and $\zeta = 1$ is the location of the maximum horizontal extent of the aquifer, i.e., $x=x_{max}$. Furthermore, the critical horizontal location $\zeta = \zeta_0$ is where the aquifer height does not change with time, i.e., $\partial h/\partial t = 0$, or where its corresponding scaled equation $ \left(1 - 2 \beta \right) \Phi + \beta \zeta \frac{\d \Phi }{\d \zeta} =  0 $ is satisfied. For dimensionless horizontal location $\zeta >\zeta_0$, the aquifer height increases with time ($\partial h/\partial t>0$), whereas for $\zeta <\zeta_0$, the aquifer height decreases with time ($\partial h/\partial t<0$).

\subsection{ Cylindrical coordinates}
In cylindrical coordinates, we implement the scaling \eqref{eq:self-similar} in the governing equations (\ref{eq:governing}-\ref{eq:condition-Kochina}) and utilize the constraint on finite volume \eqref{eq:vol} and boundary conditions at $r=0$ and $r=r_{max}$ as defined in Section \ref{sec:solutions}. The resulting ODE system for the scaled height $\Phi$ and the eigenvalue $\beta$ comes out to be

\begin{align}\label{eq:89cyl}
   \frac{\d^2 \Phi^2}{\d \zeta^2} +\frac{1}{\zeta} \frac{\d \Phi^2}{\d \zeta} + \frac{\kappa_1}{\kappa} \left[\left(1 - 2 \beta \right) \Phi + \beta \zeta \frac{\d \Phi }{\d \zeta} \right] &= 0, \quad \zeta > \zeta_0 \text{ or }\left(1 - 2 \beta \right) \Phi + \beta \zeta \frac{\d \Phi }{\d \zeta}<0\quad \text{and}\\
     \frac{\d^2 \Phi^2}{\d \zeta^2} +\frac{1}{\zeta} \frac{\d \Phi^2}{\d \zeta} +  \left(1 - 2 \beta \right) \Phi + \beta \zeta \frac{\d \Phi }{\d \zeta}  &= 0, \quad \zeta \leq \zeta_0 \text{ or }  \left(1 - 2 \beta \right) \Phi + \beta \zeta \frac{\d \Phi }{\d \zeta} \geq  0, \\
\text{subject to boundary conditions}& \hfill \nonumber\\
    \frac{\d \Phi(\zeta=0)}{\d \zeta} = 0, \quad  &  \frac{\d \Phi(\zeta=1)}{\d \zeta}  =- \frac{\kappa_1}{2\kappa} \beta \quad \text{and}\quad \Phi(\zeta=1) = 0.\label{eq:91cyl}
\end{align}
where the dimensionless location $\zeta = 0$ is the axis of the aquifer where its height is maximum, corresponding to the radial location $r=0$, and $\zeta = 1$ is the location of the maximum radial extent of the aquifer, i.e., $r=r_{max}$. Furthermore, the critical radial location $\zeta = \zeta_0$ is where the aquifer height does not change with time, i.e., $\partial h/\partial t = 0$, or where its corresponding scaled equation $ \left(1 - 2 \beta \right) \Phi + \beta \zeta \frac{\d \Phi }{\d \zeta} =  0 $ is satisfied. For dimensionless radial location $\zeta >\zeta_0$, the aquifer height increases with time ($\partial h/\partial t>0$), whereas for $\zeta <\zeta_0$, the aquifer height decreases with time ($\partial h/\partial t<0$).


\bibliographystyle{cas-model2-names}

\bibliography{jfm,afzal,cas-refs}

\end{document}